\documentclass[journal]{IEEEtran}
\usepackage[utf8]{inputenc}
\usepackage{flushend}
\usepackage{cite}
\usepackage{tabularx}
\usepackage{enumitem}
\usepackage{physics}
\usepackage{xcolor}

\ifCLASSINFOpdf
\usepackage[pdftex]{graphicx}
\else
\fi
\usepackage{amsmath,amsfonts,amsthm,amssymb}
\begin{document}
\title{Spatial Upsampling of Head-Related Transfer Functions Using a Physics-Informed 
Neural Network}
\author{Fei~Ma,~\IEEEmembership{Member,~IEEE,}
~Thushara D. Abhayapala,~\IEEEmembership{Senior Member,~IEEE},\\
~Prasanga N. Samarasinghe,~\IEEEmembership{Senior Member,~IEEE}
and~Xingyu Chen,~\IEEEmembership{Member,~IEEE}
%%%%%%%%%%%%%%%%%%%%%%%%%%%%%%%%%%%%%%%%%%%%%%%%%%%%%%%%%%%%%%%%%%%%%%%%%%%%%%%%%%%
\thanks{
Source code  https://github.com/feima1024/PINN-for-HRTF-upsampling\\
%https://github.com/AngelaYZhang/pinn-assisted-anc. \\
This work is sponsored by the Australian Research Council (ARC) 
Discovery Projects funding schemes with project numbers DP200100693.}
%%%%%%%%%%%%%%%%%%%%%%%%%%%%%%%%%%%%%%%%%%%%%%%%%%%%%%%%%%%%%%%%%%%%%%%%%%%%%%%%%%%
\thanks{Fei Ma, Thushara D. Abhayapala, Prasanga N. Samarasinghe, 
and Xingyu Chen, 
are all with the Audio and Acoustic Signal Processing Group, 
College of Engineering, Computing \& Cybernetics,
The Australian National University, Canberra, ACT 2601, Australia 
(e-mail: feima.0011@gmail.com, \{Thushara.Abhayapala, prasanga.samarasinghe,xingyu.chen1\}@anu.edu.au).}
%%%%%%%%%%%%%%%%%%%%%%%%%%%%%%%%%%%%%%%%%%%%%%%%%%%%%%%%%%%%%%%%%%%%%%%%%%%%%%%%%%%
\thanks{Manuscript received xxx xxx, xxx; revised xxx xx, xxx.}}
\maketitle

\begin{abstract}
%%%%%%%%%%%%%%%%%%%%%%%%%%%%%%%%%%%%%%%%%%%%%%%%%%%%%%%%%%%%%%%%%%%%%%%%%%
Head-related transfer function (HRTF) capture the information that a person 
uses to localize sound sources in space, and thus is crucial for creating 
personalized virtual acoustic experiences.
%%%%%%%%%%%%%%%%%%%%%%%%%%%%%%%%%%%%%%%%%%%%%%%%%%%%%%%%%%%%%%%%%%%%%%%%%%
However, practical HRTF measurement systems may only measure 
a person's HRTFs sparsely, and this necessitates HRTF upsampling. 
%%%%%%%%%%%%%%%%%%%%%%%%%%%%%%%%%%%%%%%%%%%%%%%%%%%%%%%%%%%%%%%%%%%%%%%%%%
This paper proposes a physics-informed neural network (PINN) method for
HRTF upsampling. 
%%%%%%%%%%%%%%%%%%%%%%%%%%%%%%%%%%%%%%%%%%%%%%%%%%%%%%%%%%%%%%%%%%%%%%%%%%
The PINN exploits the Helmholtz equation, the governing equation 
of acoustic wave propagation, for regularizing the upsampling process. 
%%%%%%%%%%%%%%%%%%%%%%%%%%%%%%%%%%%%%%%%%%%%%%%%%%%%%%%%%%%%%%%%%%%%%%%%%%
This helps the generation of physically valid upsamplings 
which generalize beyond the measured HRTF. 
%%%%%%%%%%%%%%%%%%%%%%%%%%%%%%%%%%%%%%%%%%%%%%%%%%%%%%%%%%%%%%%%%%%%%%%%%%
Furthermore, the size (width and depth) of the PINN is set according to 
the Helmholtz equation and its solutions, the spherical harmonics (SHs).
%%%%%%%%%%%%%%%%%%%%%%%%%%%%%%%%%%%%%%%%%%%%%%%%%%%%%%%%%%%%%%%%%%%%%%%%%%
This makes the PINN to have an appropriate level of expressive power 
and thus does not suffer from the over-fitting problem. 
%%%%%%%%%%%%%%%%%%%%%%%%%%%%%%%%%%%%%%%%%%%%%%%%%%%%%%%%%%%%%%%%%%%%%%%%%%
Since the PINN is designed independent of any specific HRTF dataset, 
it offers more generalizability compared to pure data-driven methods.
%%%%%%%%%%%%%%%%%%%%%%%%%%%%%%%%%%%%%%%%%%%%%%%%%%%%%%%%%%%%%%%%%%%%%%%%%%
Numerical experiments confirm the better performance of the PINN method 
for HRTF upsampling in both interpolation and  extrapolation scenarios 
in comparison with the SH method and the HRTF field method.
%%%%%%%%%%%%%%%%%%%%%%%%%%%%%%%%%%%%%%%%%%%%%%%%%%%%%%%%%%%%%%%%%%%%%%%%%%
\end{abstract}

%%%%%%%%%%%%%%%%%%%%%%%%%%%%%%%%%%%%%%%%%%%%%%%%%%%%%%%%%%%%%%%%%%%%%%%%%%%%%%%%%%
%%%%%%%%%%%%%%%%%%%%%%%%%%%%%%%%%%%%%%%%%%%%%%%%%%%%%%%%%%%%%%%%%%%%%%%%%%%%%%%%%%
\IEEEpeerreviewmaketitle

%%%%%%%%%%%%%%%%%%%%%%%%%%%%%%%%%%%%%%%%%%%%%%%%%%%%%%%%%%%%%%%%%%%%%%%%%%%%%%%%%%
\section{Introduction}
\label{sec:intro}
%%%%%%%%%%%%%%%%%%%%%%%%%%%%%%%%%%%%%%%%%%%%%%%%%%%%%%%%%%%%%%%%%%%%%%%%
\IEEEPARstart{H}{ead}-related transfer function (HRTF) 
is defined as the ratio between the sound pressure at a point in the ear canal 
and the sound pressure at the origin (center of head) with the head being absent~\cite{hrtf_mea}.
%%%%%%%%%%%%%%%%%%%%%%%%%%%%%%%%%%%%%%%%%%%%%%%%%%%%%%%%%%%%%%%%%%%%%%%%
HRTF characterizes the scattering effect of a person's torso, head, 
and ears with respect to the direction of sound~\cite{hrtf_mea}, 
and contains the information that a person uses to localize sound 
sources in space.
%%%%%%%%%%%%%%%%%%%%%%%%%%%%%%%%%%%%%%%%%%%%%%%%%%%%%%%%%%%%%%%%%%%%%%%%
Spatial audio and virtual acoustic systems rely on the knowledge of 
HRTF to reproduce personalized acoustic experience \cite{hrtf_personal}.  %%%%%%%%%%%%%%%%%%%%%%%%%%%%%%%%%%%%%%%%%%%%%%%%%%%%%%%%%%%%%%%%%%%%%%%%

%%%%%%%%%%%%%%%%%%%%%%%%%%%%%%%%%%%%%%%%%%%%%%%%%%%%%%%%%%%%%%%%%%%%%%%%%%%%
However, the dependence of HRTF on a person's anatomy makes HRTF highly 
individual, and thus accurate measurement of HRTF over a large number 
of directions is desirable for creating an personalized acoustic experience 
\cite{hrtf_mea}. 
%%%%%%%%%%%%%%%%%%%%%%%%%%%%%%%%%%%%%%%%%%%%%%%%%%%%%%%%%%%%%%%%%%%%%%%%%%%%
Nonetheless, a complete measurement of a person's HRTF is both time-consuming
and expensive~\cite{hrtf_mea}.
%%%%%%%%%%%%%%%%%%%%%%%%%%%%%%%%%%%%%%%%%%%%%%%%%%%%%%%%%%%%%%%%%%%%%%%%%%%%
Practical HRTF measurement systems may only conduct the measurement over a limited 
number of directions due to the inconvenience of arranging loudspeakers over 
a whole sphere \cite{hrtf_mea} or to reduce the measurement time, resulting 
in spatially sparse HRTF datasets. 
%%%%%%%%%%%%%%%%%%%%%%%%%%%%%%%%%%%%%%%%%%%%%%%%%%%%%%%%%%%%%%%%%%%%%%%%%%%%
%(Although, fast and continuous measurement systems can alleviate the time constraint \cite{time_con1,time_con2},  the high cost of these systems and% the necessary 
%anechoic chambers make them inaccessible to most people.)
%%%%%%%%%%%%%%%%%%%%%%%%%%%%%%%%%%%%%%%%%%%%%%%%%%%%%%%%%%%%%%%%%%%%%%%%%%%%
Spatially sparse HRTF datasets can compromise source localization in virtual 
acoustic environments~\cite{hrtf_vr_1,hrtf_vr_2}, prompting researchers to 
upsample them into spatially dense HRTF datasets.
%%%%%%%%%%%%%%%%%%%%%%%%%%%%%%%%%%%%%%%%%%%%%%%%%%%%%%%%%%%%%%%%%%%%%%%%%%%%

%%%%%%%%%%%%%%%%%%%%%%%%%%%%%%%%%%%%%%%%%%%%%%%%%%%%%%%%%%%%%%%%%%%%%%%%
HRTF upsampling consists of two scenarios: interpolation and 
extrapolation.
%%%%%%%%%%%%%%%%%%%%%%%%%%%%%%%%%%%%%%%%%%%%%%%%%%%%%%%%%%%%%%%%%%%%%%%%
%Note that this paper focuses on direction related HRTF 
%upsampling, and thus the 
%Distance related HRTF upsampling~\cite{hrtf_dist1,hrtf_dist2,
%hrtf_dist3,hrtf_dist4} is not addressed.
%%%%%%%%%%%%%%%%%%%%%%%%%%%%%%%%%%%%%%%%%%%%%%%%%%%%%%%%%%%%%%%%%%%%%%%
For the interpolation scenario, HRTF is measured over a limited 
number of directions to reduce the measurement time, and the aim is 
to estimate the unknown HRTF whose direction is between those of 
the measured ones.
%%%%%%%%%%%%%%%%%%%%%%%%%%%%%%%%%%%%%%%%%%%%%%%%%%%%%%%%%%%%%%%%%%%%%%%
Early works on interpolation were mainly based on the expansion of 
HRTF into some linear functions, such as  spherical harmonics 
(SHs) \cite{SH2004,SH1998,SH2012}, principle components 
\cite{PCA2012,PCA2013,PCA2020}, spline functions~\cite{Spline1999},
and wavelet functions~\cite{wavelet}.
%%%%%%%%%%%%%%%%%%%%%%%%%%%%%%%%%%%%%%%%%%%%%%%%%%%%%%%%%%%%%%%%%%%%%%
Recent works, on the other hand, are mainly based on non-linear modeling 
with neural networks (NNs) such as auto encoder~\cite{hrtf_autoencoder1,hrtf_autoencoder2,hrtf_autoencoder3},  
generative adversarial networks~\cite{hrtf_gan1,hrtf_gan2}, 
feature-wise linear modulation~\cite{film}, 
convolutional neural network~\cite{convolutional},
and  neural field~\cite{neuralfield}. 
%%%%%%%%%%%%%%%%%%%%%%%%%%%%%%%%%%%%%%%%%%%%%%%%%%%%%%%%%%%%%%%%%%%%%%%

%%%%%%%%%%%%%%%%%%%%%%%%%%%%%%%%%%%%%%%%%%%%%%%%%%%%%%%%%%%%%%%%%%%%%%%
HRTF for low evaluation angles is needed to simulate sound from 
downstairs or the sound of footsteps. 
However,  due to cost constraint, a HRTF system is of limited size and may 
not allow a convenient measurement for low evaluation angles.
%%%%%%%%%%%%%%%%%%%%%%%%%%%%%%%%%%%%%%%%%%%%%%%%%%%%%%%%%%%%%%%%%%%%%%%%%%
This results in the HRTF extrapolation scenario, and the aim is to 
estimate the unknown HRTF whose direction is beyond those of the measured ones.
%%%%%%%%%%%%%%%%%%%%%%%%%%%%%%%%%%%%%%%%%%%%%%%%%%%%%%%%%%%%%%%%%%%%%%%%%%
Although HRTF extrapolation is an essential task, it is often overlooked 
by existing research. 
Up to date, there are only a few related works. 
%%%%%%%%%%%%%%%%%%%%%%%%%%%%%%%%%%%%%%%%%%%%%%%%%%%%%%%%%%%%%%%%%%%%%%%%%%
Zhang \textit{et al.} developed iterative methods~\cite{hrtf_dir2,hrtf_dir3},  
which successively estimate the unknown HRTF for missing directions.
The methods successfully recover a low order HRTF over a full sphere with one 
quarter of data missing~\cite{hrtf_dir2}. 
%%%%%%%%%%%%%%%%%%%%%%%%%%%%%%%%%%%%%%%%%%%%%%%%%%%%%%%%%%%%%%%%%%%%%%%%%%%%%%%%%%
Duraiswami~\textit{et al.} proposed a regularized  SH method~\cite{SH2004} 
which estimates the unknown HRTF at the expense of reduced accuracy in 
representing the measured HRTF.  
%%%%%%%%%%%%%%%%%%%%%%%%%%%%%%%%%%%%%%%%%%%%%%%%%%%%%%%%%%%%%%%%%%%%%%%%%%%%%%%%%%
Ahrens \textit{et al.} proposed a non-regularized SH method~\cite{hrtf_dir4} 
which estimates the unknown HRTF based on a low-order least-square fit to the measured
HRTF and estimations of the unknown HRTF. 
%%%%%%%%%%%%%%%%%%%%%%%%%%%%%%%%%%%%%%%%%%%%%%%%%%%%%%%%%%%%%%%%%%%%%%%%%%%%%%%%%%

%%%%%%%%%%%%%%%%%%%%%%%%%%%%%%%%%%%%%%%%%%%%%%%%%%%%%%%%%%%%%%%%%%%%%%%%%%%%%%%%%%%
There are two limitations with above mentioned upsampling methods. 
%%%%%%%%%%%%%%%%%%%%%%%%%%%%%%%%%%%%%%%%%%%%%%%%%%%%%%%%%%%%%%%%%%%%%%%%%%%%%%%%%%%
First, most of the conventional linear function expansion methods, such as
SH methods~\cite{SH2004,hrtf_dir2}, had a limited exploitation of additional 
information in the upsampling process.
%%%%%%%%%%%%%%%%%%%%%%%%%%%%%%%%%%%%%%%%%%%%%%%%%%%%%%%%%%%%%%%%%%%%%%%%%%%%%%%%%%%
Their upsamplings are essentially transformations of the information that is contained 
in the measured HRTF, and thus their performance is constrained by the diversity 
of the measured HRTF. 
%%%%%%%%%%%%%%%%%%%%%%%%%%%%%%%%%%%%%%%%%%%%%%%%%%%%%%%%%%%%%%%%%%%%%%%%%%%%%%%%%%%
It was found that by exploiting the scattering field of a rigid sphere, the performance 
of SH methods can be improved~\cite{rigid}.
%%%%%%%%%%%%%%%%%%%%%%%%%%%%%%%%%%%%%%%%%%%%%%%%%%%%%%%%%%%%%%%%%%%%%%%%%%%%%%%%%%
Second, recent NN based methods~\cite{hrtf_autoencoder1,hrtf_autoencoder2,
hrtf_autoencoder3,hrtf_gan1,hrtf_gan2,film,convolutional,neuralfield}, which 
try to build up implicit associations between HRTF with additional information
(such as human anatomy and ear geometry), 
are dataset dependent~\cite{neuralfield}.  
This makes it difficult for them to extrapolate beyond the training data.
%%%%%%%%%%%%%%%%%%%%%%%%%%%%%%%%%%%%%%%%%%%%%%%%%%%%%%%%%%%%%%%%%%%%%%%%%%%%%%%%%%

%%%%%%%%%%%%%%%%%%%%%%%%%%%%%%%%%%%%%%%%%%%%%%%%%%%%%%%%%%%%%%%%%%%%%%%%%%%%%%%%%%
In recognition of these limitations, we adopt an HRTF upsampling strategy based 
on the physics-informed neural network (PINN)~\cite{pinn1, pinn2,pinn3,pinn4}.
%%%%%%%%%%%%%%%%%%%%%%%%%%%%%%%%%%%%%%%%%%%%%%%%%%%%%%%%%%%%%%%%%%%%%%%%%%%%%%%%%%
PINN is one kind of NNs which integrate physical knowledge, i.e., the 
governing partial differential equation (PDE) of a physical phenomenon, 
into its architecture~\cite{pinn1, pinn2,pinn3,pinn4}.  
The physical knowledge helps a PINN to model the physical phenomenon besides 
physical quantities. 
%%%%%%%%%%%%%%%%%%%%%%%%%%%%%%%%%%%%%%%%%%%%%%%%%%%%%%%%%%%%%%%%%%%%%%%%%%
Since the seminal works of Raissi and his colleagues \cite{pinn1,pinn2}, 
%%%%%%%%%%%%%%%%%%%%%%%%%%%%%%%%%%%%%%%%%%%%%%%%%%%%%%%%%%%%%%%%%%%%%%%%%%
PINNs have been successfully applied in many areas such as earthquake modeling~\cite{pinn_earth1,pinn_earth2}, propeller-noise prediction~\cite{pinn_drone}, 
and wave-field (sound-field) modeling \cite{pinn_4_wave_1,pinn_4_wave_2,pinn_4_wave_3,pinn_4_wave_4,pinn_4_wave_5}.     %%%%%%%%%%%%%%%%%%%%%%%%%%%%%%%%%%%%%%%%%%%%%%%%%%%%%%%%%%%%%%%%%%%%%%%%%%

%%%%%%%%%%%%%%%%%%%%%%%%%%%%%%%%%%%%%%%%%%%%%%%%%%%%%%%%%%%%%%%%%%%%%%%%%%%%%%%%%%
Owing to the principle of acoustic reciprocity~\cite{hrtf_mea}, HRTF can be 
regarded as the sound-field generated by a source placed inside of the ear canal. 
%%%%%%%%%%%%%%%%%%%%%%%%%%%%%%%%%%%%%%%%%%%%%%%%%%%%%%%%%%%%%%%%%%%%%%%%%%%%%%%%%%
This sound-field together with other sound-fields obey the Helmholtz equation, 
the governing PDE of acoustic wave propagation~\cite{william}.
%%%%%%%%%%%%%%%%%%%%%%%%%%%%%%%%%%%%%%%%%%%%%%%%%%%%%%%%%%%%%%%%%%%%%%%%%%%%%%%%%%

This fact inspires us to develop a PINN method for HRTF upsampling, and  
we inform the method with physical knowledge from two aspects. 
%%%%%%%%%%%%%%%%%%%%%%%%%%%%%%%%%%%%%%%%%%%%%%%%%%%%%%%%%%%%%%%%%%%%%%%%%%%%%%%%%%
First, a re-arranged form of the Helmholtz equation is used as the PDE loss. 
%%%%%%%%%%%%%%%%%%%%%%%%%%%%%%%%%%%%%%%%%%%%%%%%%%%%%%%%%%%%%%%%%%%%%%%%%%%%%%%%%%
This helps the PINN to generate physically valid upsamplings which generalize 
beyond the training data, and relieve the burden of balancing the PDE loss and 
the data loss with additional parameters. 
%%%%%%%%%%%%%%%%%%%%%%%%%%%%%%%%%%%%%%%%%%%%%%%%%%%%%%%%%%%%%%%%%%%%%%%%%%%%%%%%%%

Second, we set the size of the PINN method according to the SH decomposition of 
HRTF and the Helmholtz equation. 
%%%%%%%%%%%%%%%%%%%%%%%%%%%%%%%%%%%%%%%%%%%%%%%%%%%%%%%%%%%%%%%%%%%%%%%%%%
Although the PDE loss helps the PINN method to generate physically valid output, 
it also prompts the output to be zero~\cite{pinn_pp1,pinn_pp2,pinn_pp3}, 
which is a valid but trivial solution of the Helmholtz equation~\cite{pinn_pp1}.  
%%%%%%%%%%%%%%%%%%%%%%%%%%%%%%%%%%%%%%%%%%%%%%%%%%%%%%%%%%%%%%%%%%%%%%%%%%%%
This problem can be mitigated with improved gradient updating strategy~\cite{pinn_pp1,pinn_pp2}. 
Nonetheless, the strategy~\cite{pinn_pp1,pinn_pp2} is computationally 
expensive and requires expert knowledge to tune additional parameters. 
%%%%%%%%%%%%%%%%%%%%%%%%%%%%%%%%%%%%%%%%%%%%%%%%%%%%%%%%%%%%%%%%%%%%%%%%%%%%
We found that this problem is due to over-fitting, i.e., over-parameterization 
of PINN methods. 
%%%%%%%%%%%%%%%%%%%%%%%%%%%%%%%%%%%%%%%%%%%%%%%%%%%%%%%%%%%%%%%%%%%%%%%%%%%%
On recognition of the over-fitting, we design the PINN method with an appropriate 
level of expressive power by exploiting the solution of the Helmholtz equation 
in spherical coordinates, the SHs. 
%%%%%%%%%%%%%%%%%%%%%%%%%%%%%%%%%%%%%%%%%%%%%%%%%%%%%%%%%%%%%%%%%%%%%%%%%%%%
Specifically, we set the PINN method width (the number of neurons in each hidden layers) 
as half of the dimensionality of HRTFs under SH decomposition \cite{hrtf_insight,Thushara_kr,width}, 
and the depth of it (the number of hidden layers) as three. 
%%%%%%%%%%%%%%%%%%%%%%%%%%%%%%%%%%%%%%%%%%%%%%%%%%%%%%%%%%%%%%%%%%%%%%%%%%%%%%%%%%
This setup separates the proposed PINN method apart from general PINN methods 
in other works that suffer from the over-fitting problem 
due to inappropriate design of the network~\cite{pinn_pp1,pinn_pp2,pinn_pp3}. 
%%%%%%%%%%%%%%%%%%%%%%%%%%%%%%%%%%%%%%%%%%%%%%%%%%%%%%%%%%%%%%%%%%%%%%%%%%%%

The effective exploitation of the data-independent Helmholtz equation and the SHs
compensates for the lack of measured data, and grants the PINN method 
with extrapolation ability.
%%%%%%%%%%%%%%%%%%%%%%%%%%%%%%%%%%%%%%%%%%%%%%%%%%%%%%%%%%%%%%%%%%%%%%%%%%%%%%%%%%
The performance of the PINN method for upsampling HRTF in both interpolation and 
extrapolation scenarios are confirmed by numerical experiments, and compared with 
the SH method~\cite{SH2004} and the HRTF field method~\cite{neuralfield}.
%%%%%%%%%%%%%%%%%%%%%%%%%%%%%%%%%%%%%%%%%%%%%%%%%%%%%%%%%%%%%%%%%%%%%%%%%%%%%%%%%%%

%%%%%%%%%%%%%%%%%%%%%%%%%%%%%%%%%%%%%%%%%%%%%%%%%%%%%%%%%%%%%%%%%%%%%%%%%%%%%%%%%%%
The rest of this paper is organized as follows. 
The problem of interest is introduced in Sec.~II.
%%%%%%%%%%%%%%%%%%%%%%%%%%%%%%%%%%%%%%%%%%%%%%%%%%%%%%%%%%%%%%%%%%%%%%%%%%%%%%%%%%%
We review the SH method~\cite{SH2004} in Sec.~III 
and propose a PINN method in Sec.~IV. 
%%%%%%%%%%%%%%%%%%%%%%%%%%%%%%%%%%%%%%%%%%%%%%%%%%%%%%%%%%%%%%%%%%%%%%%%%%%%%%%%%%%
In Secs. V and VI, we compare the performance of the PINN method, 
the SH method~\cite{SH2004}, the standard NN method, and the HRTF 
field method~\cite{neuralfield} 
using interpolation and extrapolation experiments, respectively.
%%%%%%%%%%%%%%%%%%%%%%%%%%%%%%%%%%%%%%%%%%%%%%%%%%%%%%%%%%%%%%%%%%%%%%%%%%%%%%%%%%%
Section VII discusses the experiment results, 
points out directions of improvement,
and presents limitations of the PINN method. 
Section VIII concludes this paper.
%%%%%%%%%%%%%%%%%%%%%%%%%%%%%%%%%%%%%%%%%%%%%%%%%%%%%%%%%%%%%%%%%%%%%%%%%%%%%%%%%%%%%

%%%%%%%%%%%%%%%%%%%%%%%%%%%%%%%%%%%%%%%%%%%%%%%%%%%%%%%%%%%%%%%%%%%%%%%%%%
\begin{figure}[t]
\centering
\includegraphics[trim={0cm 0cm 0 0.cm},clip,width=8.0cm]{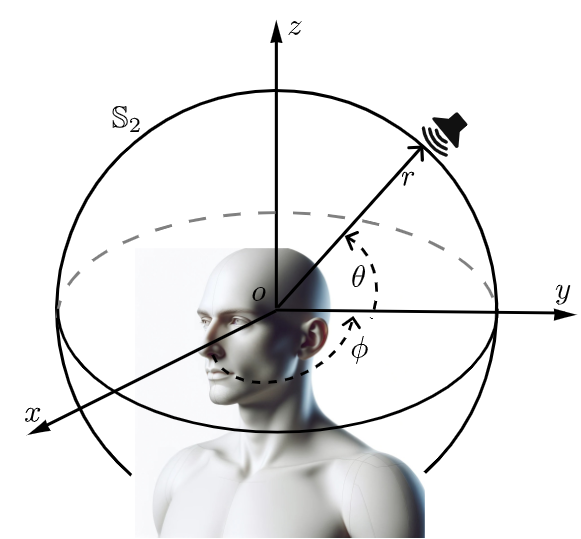}
\caption{Layout of a typical HRTF measurement system, which measures the 
HRTF between a loudspeaker placed on the sphere $\mathbb{S}_2$ and a 
microphones placed at a specific position inside of the person's ear, 
ideally close to the ear drum.
A Cartesian coordinate system and a spherical coordinate system are set up with 
respect to center of the person's head. 
The person is facing the positive $x$-axis. 
HRTF is measured from various directions by rotating either the 
loudspeaker or the person.}
\label{fig:measurement}
\end{figure}
%%%%%%%%%%%%%%%%%%%%%%%%%%%%%%%%%%%%%%%%%%%%%%%%%%%%%%%%%%%%%%%%%%%%%%%%%%

%%%%%%%%%%%%%%%%%%%%%%%%%%%%%%%%%%%%%%%%%%%%%%%%%%%%%%%%%%%%%%%%%%%%%%%%%%
\section{Problem formulation}
\label{sec:problem}
%%%%%%%%%%%%%%%%%%%%%%%%%%%%%%%%%%%%%%%%%%%%%%%%%%%%%%%%%%%%%%%%%%%%%%%%%%%%%%%%%%%%%%
Figure \ref{fig:measurement} presents the layout of a typical HRTF measurement system~\cite{hrtf_mea}. 
%%%%%%%%%%%%%%%%%%%%%%%%%%%%%%%%%%%%%%%%%%%%%%%%%%%%%%%%%%%%%%%%%%%%%%%%%%%%%%%%%%%%%%
Let $(x,y,z)$ and $(r,\theta,\phi)$ denote the Cartesian coordinates and the 
spherical coordinates of a point with respect to the center of a person's ears.
%%%%%%%%%%%%%%%%%%%%%%%%%%%%%%%%%%%%%%%%%%%%%%%%%%%%%%%%%%%%%%%%%%%%%%%%%%%%%%%%%%%%%%
%The system measures the HRTF between a loudspeaker placed on the sphere 
%$\mathbb{S}_2$ and a microphone placed at a specific position in the ear canal, 
%ideally close to the ear drum. 
%The HRTF is measured from various directions by rotating either the loudspeaker or the person.
%%%%%%%%%%%%%%%%%%%%%%%%%%%%%%%%%%%%%%%%%%%%%%%%%%%%%%%%%%%%%%%%%%%%%%%%%%%
We denote HRTF as $P(\omega,r,\theta,\phi)$ in spherical coordinates or 
as $P(\omega,x,y,z)$ in Cartesian coordinates, where   
$\omega=2\pi{f}$ is the angular frequency and $f$ is the frequency. 
%%%%%%%%%%%%%%%%%%%%%%%%%%%%%%%%%%%%%%%%%%%%%%%%%%%%%%%%%%%%%%%%%%%%%%%%%%%
Hereafter, HRTF is evaluated on a single sphere, and thus we skip the 
sphere radius $r$ when representing HRTF and related acoustic quantities 
for notational simplicity. 
%%%%%%%%%%%%%%%%%%%%%%%%%%%%%%%%%%%%%%%%%%%%%%%%%%%%%%%%%%%%%%%%%%%%%%%%%%%

%%%%%%%%%%%%%%%%%%%%%%%%%%%%%%%%%%%%%%%%%%%%%%%%%%%%%%%%%%%%%%%%%%%%%%%%%%%%%%%%%%%
As shown in Fig.~\ref{fig:measurement}, 
the design of the system does not allow a convenient full sphere HRTF measurement. 
To reduce the measurement time, the system  may only measure HRTF 
over a limited number of directions.
%%%%%%%%%%%%%%%%%%%%%%%%%%%%%%%%%%%%%%%%%%%%%%%%%%%%%%%%%%%%%%%%%%%%%%%%%%%%%%%%%%%
Both of these two scenarios will result in spatially sparse HRTF datasets, 
which may be insufficient for virtual acoustic applications~\cite{hrtf_vr_1,hrtf_vr_2}.
%%%%%%%%%%%%%%%%%%%%%%%%%%%%%%%%%%%%%%%%%%%%%%%%%%%%%%%%%%%%%%%%%%%%%%%%%%%%%%%%%%%

Sound-fields obey the governing PDE of acoustic 
wave propagation, i.e., the Helmholtz equation,~\cite{william}
\begin{IEEEeqnarray}{rcl}
\label{eq:wave0}
\nabla^2P+(w/c)^2P=0,
\end{IEEEeqnarray} 
where $c$ is the speed of sound, and $\nabla^2$ denotes the Laplacian operator. 
%%%%%%%%%%%%%%%%%%%%%%%%%%%%%%%%%%%%%%%%%%%%%%%%%%%%%%%%%%%%%%%%%%%%%%%%%%%%%%%%%%%%%%%%%%%%%
In spherical coordinates, the Laplacian is given by~\cite{william}
\begin{IEEEeqnarray}{rcl}
\label{eq:wave1}
\nabla^2P&=&
\frac{2}{r}
\frac{\partial{}P}{\partial{}r}+
\frac{\partial{}^2P}{\partial{}r^2}+				
\frac{\cos\theta}{r^2\sin{\theta}}
\frac{\partial{P}}{\partial{\theta}}        \nonumber\\
&&+
\frac{1}{r^2}
\frac{\partial^2{P}}{\partial{\theta}^2}    
+\frac{1}{r^2\sin^2{\theta}} 
\frac{\partial^2{P}}{\partial{}\phi^2},
\end{IEEEeqnarray} 
%%%%%%%%%%%%%%%%%%%%%%%%%%%%%%%%%%%%%%%%%%%%%%%%%%%%%%%%%%%%%%%%%%%%%%%%%%%%%%%%%%%%%%%%%%%%%
and in Cartesian coordinates it is given by
\begin{IEEEeqnarray}{rcl}
\label{eq:wave2}
\nabla^2P=
 \frac{\partial^2{P}}{\partial{x^2}} 
+\frac{\partial^2{P}}{\partial{y^2}} 
+\frac{\partial^2{P}}{\partial{z^2}}. 
\end{IEEEeqnarray} 
%%%%%%%%%%%%%%%%%%%%%%%%%%%%%%%%%%%%%%%%%%%%%%%%%%%%%%%%%%%%%%%%%%%%%%%%%%%%%%%%%%%%%%%%%%%%%

%%%%%%%%%%%%%%%%%%%%%%%%%%%%%%%%%%%%%%%%%%%%%%%%%%%%%%%%%%%%%%%%%%%%%%%%%%%
In this paper, by exploiting the Helmholtz equation Eq.~\eqref{eq:wave0}  
and its solution in spherical coordinates, the SHs, we aim to upsample 
a spatially sparse HRTF dataset $\{P(\omega,\theta_q,\phi_q)\}_{q=1}^{Q}$ 
or equivalently $\{P(\omega,x_q,y_q,z_q)\}_{q=1}^{Q}$ into a spatially 
dense HRTF dataset. 
%%%%%%%%%%%%%%%%%%%%%%%%%%%%%%%%%%%%%%%%%%%%%%%%%%%%%%%%%%%%%%%%%%%%%%%%%%%
$Q$ is the number of sampling points and $q$ is the index of a particular 
sampling point.
%%%%%%%%%%%%%%%%%%%%%%%%%%%%%%%%%%%%%%%%%%%%%%%%%%%%%%%%%%%%%%%%%%%%%%%%%%%

%%%%%%%%%%%%%%%%%%%%%%%%%%%%%%%%%%%%%%%%%%%%%%%%%%%%%%%%%%%%%%%%%%%%%%%%%%%
%\section{THEORETICAL BACKGROUND}
\section{Spherical-harmonics-based method}
\label{sec:sh}
In this section, we first briefly present the SH decomposition of 
HRTF and then review the regularized SH method~\cite{SH2004} 
for HRTF upsampling. 
%%%%%%%%%%%%%%%%%%%%%%%%%%%%%%%%%%%%%%%%%%%%%%%%%%%%%%%%%%%%%%%%%%%%%%%%%%%
HRTF is expressed in spherical coordinates for the ease of SH 
decomposition. 
%%%%%%%%%%%%%%%%%%%%%%%%%%%%%%%%%%%%%%%%%%%%%%%%%%%%%%%%%%%%%%%%%%%%%%%%%%%

%%%%%%%%%%%%%%%%%%%%%%%%%%%%%%%%%%%%%%%%%%%%%%%%%in spherical coordinates
HRTF can be decomposed into SHs %and their coefficients 
as~\cite{william} 
\begin{IEEEeqnarray}{rcl}
\label{eq:SH_first}
\mathbf{P}=\mathbf{Y}\mathbf{A}, 
\end{IEEEeqnarray}
where 
$\mathbf{P}=[ 
P(\omega,\theta_1,\phi_1),  
P(\omega,\theta_2,\phi_2),  
...,  
P(\omega,\theta_Q,\phi_Q) ]^{\intercal}$ denote the measured 
HRTF for directions $(\theta_q,\phi_q)_{q=1}^{Q}$,
$(\cdot)^\intercal$ is the transpose operation,
$\mathbf{A}=[ 
A_{0,0}(\omega),      
A_{1,-1}(\omega), 
..., 
A_{U,U}(\omega)
]^{\intercal} $ 
denote the SH coefficients, 
and 
\begin{IEEEeqnarray}{rcl}
\mathbf{Y}=
\begin{bmatrix}
\label{eq:sh_second}
Y_{0}^{0}(\theta_1,\phi_1) & 
Y_{1}^{-1}(\theta_1,\phi_1) & 
... &
Y_{U}^{U}(\theta_1,\phi_1)  \\
%%%%%%%%%%%%%%%%%%%%%%%%%%%%%%%
Y_{0}^{0}(\theta_2,\phi_2) & 
Y_{1}^{-1}(\theta_2,\phi_2) & 
...&
Y_{U}^{U}(\theta_2,\phi_2)  \\
%%%%%%%%%%%%%%%%%%%%%%%%%%%%%%%
... & 
... & 
... &
... \\
%%%%%%%%%%%%%%%%%%%%%%%%%%%%%%%
Y_{0}^{0}(\theta_Q,\phi_Q) & 
Y_{1}^{-1}(\theta_Q,\phi_Q) & 
... &
Y_{U}^{U}(\theta_Q,\phi_Q)
\end{bmatrix},\qquad
\end{IEEEeqnarray}
denotes a $Q\times(U+1)^2$ matrix whose entries are SHs $Y_{u}^{v}(\cdot,\cdot)$ 
of order $u$ and  degree $v$. 
SHs are defined as~\cite{hutubs} 
\begin{IEEEeqnarray}{rCl}
\label{eq:rsh}
Y_{u}^{v}(\theta,\phi)
&\equiv&
\sqrt{\frac{(2u+1)(u-|v|)!}{4\pi(u+|v|)!}}
\mathcal{P}_{u}^{|v|}(\sin\theta)e^{iv\phi},
\end{IEEEeqnarray}  
%%%%%%%%%%%%%%%%%%%%%%%%%%%%%%%%%%%%%%%%%%%%%%%%%%%%%%%%%%%%%%%%%%%%%%%%%%
where $|\cdot|$ is the absolute value operator, $\mathcal{P}_{u}^{|v|}(\cdot)$ 
is the associated Legendre function of order $u$ and  degree $|v|$, 
$i=\sqrt{-1}$ is the imaginary unit, and $e^{(\cdot)}$ is the exponential function.
%%%%%%%%%%%%%%%%%%%%%%%%%%%%%%%%%%%%%%%%%%%%%%%%%%%%%%%%%%%%%%%%%%%%%%%%%%
SHs is the solution of the Helmholtz equation for the elevation angle $\theta$ 
and the azimuth angle $\phi$~\cite{william}. 
%%%%%%%%%%%%%%%%%%%%%%%%%%%%%%%%%%%%%%%%%%%%%%%%%%%%%%%%%%%%%%%%%%%%%%%%%%

\begin{figure}[t]
\centering
\includegraphics[width=7.8cm]{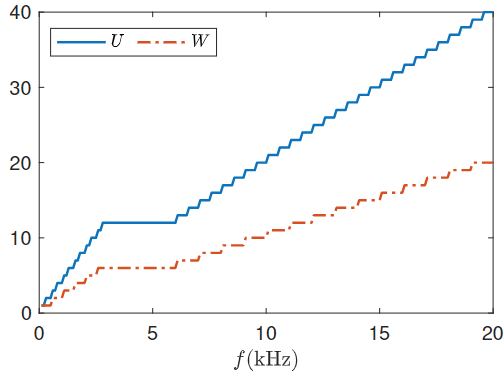}
\caption{Dimensionality $U$ of HRTFs under SH decomposition and 
the PINN width $W$ as functions of frequency $f$.}
\label{fig:dimensionality}
\end{figure}

%%%%%%%%%%%%%%%%%%%%%%%%%%%%%%%%%%%%%%%%%%%%%%%%%%%%%%%%%%%%%%%%%%%%%%%%%%%%%%%%
In Eqs.~\eqref{eq:SH_first} and \eqref{eq:sh_second}, 
$U$ is the dimensionality of HRTF under SH decomposition 
and can be approximated as~\cite{hrtf_insight,Thushara_kr} 
\begin{IEEEeqnarray}{rcl}
\label{eq:order}
U=\lceil{2\pi{f}r_\mathrm{h}/c}\rceil,
\end{IEEEeqnarray}
where $\lceil{\cdot}\rceil$ is the ceiling operation, 
and 
\begin{IEEEeqnarray}{rcl}
r_\mathrm{h}=
\begin{cases}
0.2 \;\mathrm{m}, & f\leq3\; \mathrm{kHz},\\ 
0.09 \;\mathrm{m}, & f>3\; \mathrm{kHz}, 
\end{cases}
\end{IEEEeqnarray}
is the radius of 
human head, including the head-and-torso scattering effect.
%%%%%%%%%%%%%%%%%%%%%%%%%%%%%%%%%%%%%%%%%%%%%%%%%%%%%%%%%%%%%%%%%%%%%%%%%%%%%%%%
In this paper, for simplicity, we approximate Eq.~\eqref{eq:order} as
%%%%%%%%%%%%%%%%%%%%%%%%%%%%%%%%%%%%%%%%%%%%%%%%%%%%%%%%%%%%%%%%%%%%%%%%%%%%%%%%
\begin{IEEEeqnarray}{rcl}
\label{eq:size}
U%&=&\lceil{2\pi{f}r_\mathrm{h}/c}\rceil       \nonumber\\
&\approx&
\begin{cases}
\lceil{f/250}\rceil, & f<3 \; \mathrm{kHz},\\ 
12, & 3 \;\mathrm{kHz} \leq f\leq6 \;\mathrm{kHz},\\ 
\lceil{f/500}\rceil, & f > 6 \;\mathrm{kHz}, 
\end{cases}
\end{IEEEeqnarray}
and present $U$ as a function of frequency in Fig.~2 for reference.
%%%%%%%%%%%%%%%%%%%%%%%%%%%%%%%%%%%%%%%%%%%%%%%%%%%%%%%%%%%%%%%%%%%%%%%%%%%%%%%%
Since the dimensions of human heads are highly individual, 
Eq.~\eqref{eq:size} and Fig.~\ref{fig:dimensionality} represent approximations only.
%%%%%%%%%%%%%%%%%%%%%%%%%%%%%%%%%%%%%%%%%%%%%%%%%%%%%%%%%%%%%%%%%%%%%%%%%%%%%%%%

%%%%%%%%%%%%%%%%%%%%%%%%%%%%%%%%%%%%%%%%%%%%%%%%%%%%%%%%%%%%%%%%%%%%%%%%%
The regularized SH method~\cite{SH2004} first estimates the SH coefficients 
$\hat{\mathbf{A}}=[ 
\hat{A}_{0,0}(\omega),      
\hat{A}_{1,-1}(\omega), 
..., 
\hat{A}_{U,U}(\omega)
]^{\intercal} $ 
through
\begin{IEEEeqnarray}{rcl}
\label{eq:regulation}
\hat{\mathbf{A}}=(\mathbf{Y}^{H}\mathbf{Y}+\gamma\mathbf{H})^{-1}\mathbf{Y}^{H}
\mathbf{P}, 
\end{IEEEeqnarray}
%%%%%%%%%%%%%%%%%%%%%%%%%%%%%%%%%%%%%%%%%%%%%%%%%%%%%%%%%%%%%%%%%%%%%%%%%
where $(\cdot)^{H}$ denotes the complex-conjugate operation, $\mathbf{H}$ is a $(U+1)^2\times(U+1)^2$ diagonal matrix whose diagonal 
entries are $h_{l,l}=1+u(u+1)$ and $\gamma$ is the regularization parameter. 
The regularization limits the estimated SH coefficients  $\hat{\mathbf{A}}$,
especially the high-order coefficients, from taking large values~\cite{SH2004}.
%%%%%%%%%%%%%%%%%%%%%%%%%%%%%%%%%%%%%%%%%%%%%%%%%%%%%%%%%%%%%%%%%%%%%%%%%
To estimate the SH coefficients up to order $U$, the number of measured HRTFs 
needs to be sufficiently large, i.e., $Q>(U+1)^2$~\cite{hrtf_insight,Thushara_kr}. 
%%%%%%%%%%%%%%%%%%%%%%%%%%%%%%%%%%%%%%%%%%%%%%%%%%%%%%%%%%%%%%%%%%%%%%%%%%%%

%%%%%%%%%%%%%%%%%%%%%%%%%%%%%%%%%%%%%%%%%%%%%%%%%%%%%%%%%%%%%%%%%%%%%%%%%%%%
The HRTF for an arbitrary direction $(\theta_e,\phi_e)$ can be estimated as
\begin{IEEEeqnarray}{rcl}
\label{eq:sh_last}
\hat{P}_{\mathrm{SH}}(\omega,\theta_e,\phi_e)
&\approx&\mathbf{Y}_{e}\hat{\mathbf{A}},
\end{IEEEeqnarray}
%%%%%%%%%%%%%%%%%%%%%%%%%%%%%%%%%%%%%%%%%%%%%%%%%%%%%%%%%%%%%%%%%%%%%%%%%%%%
where $\mathbf{Y}_e=[Y_{0}^{0}(\theta_e,\phi_e), Y_{1}^{-1}(\theta_e,\phi_e),  
..., Y_{U}^{U}(\theta_e,\phi_e)]$.
%%%%%%%%%%%%%%%%%%%%%%%%%%%%%%%%%%%%%%%%%%%%%%%%%%%%%%%%%%%%%%%%%%%%%%%%%%%%

%%%%%%%%%%%%%%%%%%%%%%%%%%%%%%%%%%%%%%%%%%%%%%%%%%%%%%%%%%%%%%%%%%%%%%%%%%%%
\begin{figure*}[t]
\centering
\includegraphics[width=16cm]{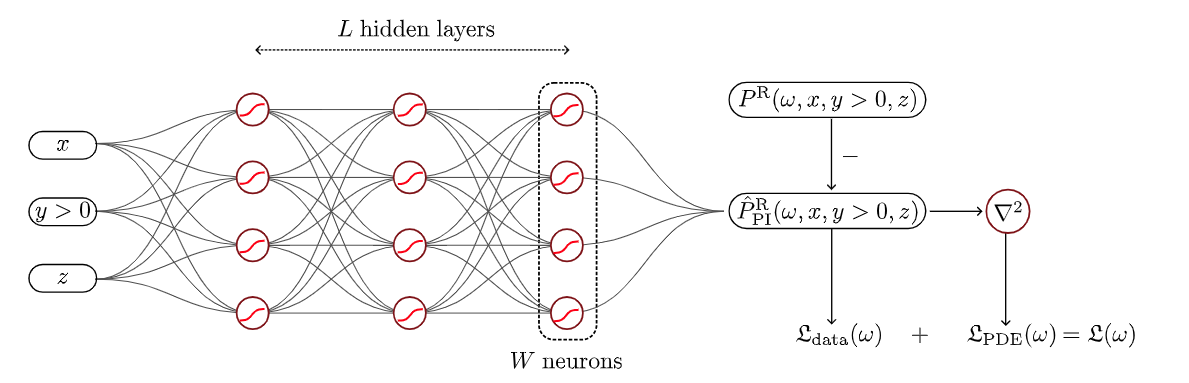}
\caption{Structure of the proposed PINN method for modeling the real and left 
part HRTF: The inputs are Cartesian coordinates $(x, y>0, z)$ and the output 
is HRTF estimation $\hat{P}_{\mathrm{PI}}^{\mathrm{R}}(\omega, x, y>0, z)$. 
There are $L$ hidden layers with $W$ neurons in each hidden layer, and  the
activation function is $\tanh$. 
Data loss and PDE loss are calculated with respect to HRTF estimation 
$\hat{P}_{\mathrm{PI}}^{\mathrm{R}}(\omega, x, y>0, z)$ and its Laplacian 
$\nabla^2$, respectively.}
\label{fig:pinn}
\end{figure*}
%%%%%%%%%%%%%%%%%%%%%%%%%%%%%%%%%%%%%%%%%%%%%%%%%%%%%%%%%%%%%%%%%%%%%%%%%%%%

%%%%%%%%%%%%%%%%%%%%%%%%%%%%%%%%%%%%%%%%%%%%%%%%%%%%%%%%%%%%%%%%%%%%%%%%%%%%
\section{Physics-informed-neural-network-based method}\label{sec:nn}
In this section, we first briefly introduce general PINN methods, 
then propose a PINN method for HRTF upsampling, and at 
last provide rationals for the configuration of the proposed 
PINN method.

\subsection{General PINN methods}\label{sec:pinn}
%%%%%%%%%%%%%%%%%%%%%%%%%%%%%%%%%%%%%%%%%%%%%%%%%%%%%%%%%%%%%%%%%%%%%%%%%%%%
PINN methods are commonly constructed as multi-layer fully connected feed-forward 
neural networks~\cite{pinn1,pinn2,pinn3,pinn4}. 
%%%%%%%%%%%%%%%%%%%%%%%%%%%%%%%%%%%%%%%%%%%%%%%%%%%%%%%%%%%%%%%%%%%%%%%%%%%%
The functionality of one layer is
\begin{IEEEeqnarray}{rcl}
\mathcal{P}(\mathbf{x})=
[\sigma\left(\mathbf{x}^{\intercal} \mathbf{w}_1+b_1\right),
\sigma\left(\mathbf{x}^{\intercal} \mathbf{w}_2+b_2\right),
..., 
\sigma\left(\mathbf{x}^{\intercal} \mathbf{w}_J+b_J\right)]^{\intercal}, \nonumber\\
\end{IEEEeqnarray}
where $\mathbf{x}$ is the input variable vector, $\{\mathbf{w}_{j=1}^{J}\}$ 
are the weight vectors, $\{b_j\}_{j=1}^{J}$ are the biases, 
$J$ is the number of neurons, and $\sigma(\cdot)$ is the activation function. 
%%%%%%%%%%%%%%%%%%%%%%%%%%%%%%%%%%%%%%%%%%%%%%%%%%%%%%%%%%%%%%%%%%%%%%%%%%%%
The overall functionality of a PINN method is the composition of $L$ hidden 
layers and one output layer
\begin{equation}
\hat{P}(\mathbf{x} ;{\zeta})
=\mathbf{w}_o
\mathcal{P}_L( .... (\mathcal{P}_2  (\mathcal{P}_1(\mathbf{x})))) + b_o,
\end{equation}
where $\mathbf{w}_o$ and $b_o$ denote the output layer weight vector and 
basis, respectively,
$\zeta$ represents the set of trainable parameters,
and $\hat{P}(\mathbf{x} ;{\zeta})$ represents the predicted sound pressure. 
$\zeta$ is adjusted by minimizing a loss function 
\begin{IEEEeqnarray}{rcl}
\label{eq:conventionalloss}
\mathfrak{L}&=&(1-\gamma)
\frac{1}{Q}\sum_{q=1}^Q\left(P_q-\hat{P}(\mathbf{x}_q;\zeta)\right)^2 
%\nonumber\\&&
+\lambda\mathfrak{L}_\mathrm{PDE}(\mathbf{x};\zeta), 
%, \mathbf{x}\in{\Omega},
\end{IEEEeqnarray}
%%%%%%%%%%%%%%%%%%%%%%%%%%%%%%%%%%%%%%%%%%%%%%%%%%%%%%%%%%%%%%%%%%%%%%%%%%%%
where $\{\mathbf{x}_q,P_q\}_{q=1}^{Q}$ are input-output training data 
pairs which are obtained by testing and measuring a physical system, 
$\mathfrak{L}_\mathrm{PDE}(\mathbf{x};\zeta)$ corresponds to the residual 
of the governing PDE, and $\lambda\geq0$ is a regularization parameter which 
balances the contributions of two loss terms to the total loss $\mathfrak{L}$. 
%%%%%%%%%%%%%%%%%%%%%%%%%%%%%%%%%%%%%%%%%%%%%%%%%%%%%%%%%%%%%%%%%%%%%%%%%%%%

%%%%%%%%%%%%%%%%%%%%%%%%%%%%%%%%%%%%%%%%%%%%%%%%%%%%%%%%%%%%%%%%%%%%%%%%%%%%
\subsection{Proposed PINN method\label{sec:proposed}}

%%%%%%%%%%%%%%%%%%%%%%%%%%%%%%%%%%%%%%%%%%%%%%%%%%%%%%%%%%%%%%%%%%%%%%%%%%%%
We use four PINN methods to model the HRTF for one ear of a person 
at a single frequency $\omega$.
%%%%%%%%%%%%%%%%%%%%%%%%%%%%%%%%%%%%%%%%%%%%%%%%%%%%%%%%%%%%%%%%%%%%%%%%%%%%%%%%%
Specifically, for  
\begin{enumerate}
\item the real and left part ${{P}^{\mathrm{R}}(\omega,x,y>0,z)}$; 
\item the real and right part  ${P^{\mathrm{R}}(\omega,x,y<0,z)}$;
\item the imaginary and left  part ${P^{\mathrm{I}}(\omega,x,y>0,z)}$; 
\item the imaginary and right part  ${P^{\mathrm{I}}(\omega,x,y<0,z)}$. 
\end{enumerate}
%%%%%%%%%%%%%%%%%%%%%%%%%%%%%%%%%%%%%%%%%%%%%%%%%%%%%%%%%%%%%%%%%%%%%%%%%%%%%%%%%
The superscripts ${(\cdot)}^{\mathrm{R}}$ and ${(\cdot)}^{\mathrm{I}}$ 
denote the real part and the imaginary part of a value, respectively.
%%%%%%%%%%%%%%%%%%%%%%%%%%%%%%%%%%%%%%%%%%%%%%%%%%%%%%%%%%%%%%%%%%%%%%%%%%%%%%%%%
With the person facing the positive $x$-axis as shown in Fig.~\ref{fig:measurement}, 
$y>0$ and $y<0$ denote the left side HRTF and the right side HRTF, respectively. 
%%%%%%%%%%%%%%%%%%%%%%%%%%%%%%%%%%%%%%%%%%%%%%%%%%%%%%%%%%%%%%%%
This is made for two reasons.  
%%%%%%%%%%%%%%%%%%%%%%%%%%%%%%%%%%%%%%%%%%%%%%%%%%%%%%%%%%%%%%%%%%%%%%%%%%%%
First, when the loudspeaker is at the same side with 
the ear the magnitude of HRTF tends to be larger than the 
magnitude of HRTF when the loudspeaker is at the opposite side to the ear.
%%%%%%%%%%%%%%%%%%%%%%%%%%%%%%%%%%%%%%%%%%%%%%%%%%%%%%%%%%%%%%%%%%%%%%%%%%%%
The magnitude difference subsequently affects their contributions to the loss function, 
making a single PINN method difficult to attain consistent upsampling accuracy for both sides. 
Second, training real-valued neural networks is simpler compared 
to their complex-valued counterparts~\cite{complex}. 
%%%%%%%%%%%%%%%%%%%%%%%%%%%%%%%%%%%%%%%%%%%%%%%%%%%%%%%%%%%%%%%%%%%%%%%%%%%%

%%%%%%%%%%%%%%%%%%%%%%%%%%%%%%%%%%%%%%%%%%%%%%%%%%%%%%%%%%%%%%%%%%%%%%%%%%%%%%%%%
The structure of the PINN method for modeling the real and left part 
HRTF ${P^{\mathrm{R}}(\omega,x,y>0,z)}$ is shown in Fig.~\ref{fig:pinn}.

%%%%%%%%%%%%%%%%%%%%%%%%%%%%%%%%%%%%%%%%%%%%%%%%%%%%%%%%%%%%%%%%%%%%%%%%%%%%%%%%%
%There are $L$ hidden layers with $W$ neurons in each hidden layer. 
%The inputs are Cartesian coordinates $(x>0, y, z)$, the activation 
%functions are $\tanh$, and the output is HRTF estimation 
%$\hat{P}_{\mathrm{PI}}^{\mathrm{R}}(\omega,x>0,y,z)$. 
%%%%%%%%%%%%%%%%%%%%%%%%%%%%%%%%%%%%%%%%%%%%%%%%%%%%%%%%%%%%%%%%%%%%%%%%%%%%%%%%%
%%%%%%%%%%%%%%%%%%%%%%%%%%%%%%%%%%%%%%%%%%%%%%%%%%%%%%%%%%%%%%%%%%%%%%%%%%%%
The loss function for this specific case is given by
\begin{IEEEeqnarray}{rcl}
\label{eq:cost}
&&\mathfrak{L}(\omega)= 
\underbrace{
\frac{1}{Q}\sum_{q=1}^{Q}(
P^{\mathrm{R}}(\omega,x_q,y_q>0,z_q)-
\hat{P}_{\mathrm{PI}}^{\mathrm{R}}(\omega,x_q,y_q>0,z_q)
)^2
}_{\mathfrak{L}_{\mathrm{data}}(\omega)}
\nonumber\\
&&+
\underbrace{
\frac{1}{D}
\sum_{d=1}^{D}
( \frac{
\nabla^2{}\hat{P}_{\mathrm{PI}}^{\mathrm{R}}(\omega,x_d,y_d>0,z_d) 
}{(w/c)^2}
%\nabla^2{}\hat{P}_{\mathrm{PI}}(\omega,x_d,y_d,z_d) 
+\hat{P}_{\mathrm{PI}}^{\mathrm{R}}(\omega,x_d,y_d>0,z_d) )^2
}_{\mathfrak{L}_{\mathrm{PDE}}(\omega)}, \nonumber\\
\end{IEEEeqnarray}
where the Laplacian operator $\nabla^2$ is given by Eq.~\eqref{eq:wave2},
%\equiv \frac{\partial^2}{\partial{}x^2} 
%+ \frac{\partial^2}{\partial{}y^2}    
%+ \frac{\partial^2}{\partial{}z^2}$ 
%is the Laplacian operator~\cite{william},
$\{x_q, y_q>0, z_q\}_{q=1}^{Q}$ are Cartesian coordinates of the measured HRTF 
$P^{\mathrm{R}}(\omega,x_q,y_q>0,z_q)$, 
$\{x_d, y_d>0, z_d\}_{d=1}^{D}$ denote the Cartesian coordinates (or directions) 
of the unknown HRTF we want to upsample and is a super set of $\{(x_q, y_q>0, z_q)\}_{q=1}^{Q}$, 
and $\mathfrak{L}_{\mathrm{data}}(\omega)$ and $\mathfrak{L}_{\mathrm{PDE}}(\omega)$
denote data loss and PDE loss, respectively. 
%%%%%%%%%%%%%%%%%%%%%%%%%%%%%%%%%%%%%%%%%%%%%%%%%%%%%%%%%%%%%%%%%%%%%%%%%%%%
There is no regularization parameter $\lambda$ in Eq.~\eqref{eq:cost} for balancing the data loss
and the PDE loss, and the absence of $\lambda$ is explained in
Sec.~\ref{sec:pinn} C.
%%%%%%%%%%%%%%%%%%%%%%%%%%%%%%%%%%%%%%%%%%%%%%%%%%%%%%%%%%%%%%%%%%%%%%%%%%%%
Note that we regard HRTF as a sound-field around a human head and thus
the Cartesian coordinates in~Eq.~\eqref{eq:cost} correspond to 
$(r_\mathrm{h},\theta, 0\leq\phi<\pi)$. 
%%%%%%%%%%%%%%%%%%%%%%%%%%%%%%%%%%%%%%%%%%%%%%%%%%%%%%%%%%%%%%%%%%%%%%%%%%%%

Except the training data and the output, the structures of PINN methods for modeling other three 
parts and the loss functions are identical to Fig.~\ref{fig:pinn} and Eq.~\eqref{eq:cost}, 
respectively.  
%%%%%%%%%%%%%%%%%%%%%%%%%%%%%%%%%%%%%%%%%%%%%%%%%%%%%%%%%%%%%%%%%%%%%%%%%%%%
Once trained, we combine the outputs of four PINN methods to arrive at the complex 
value HRTF for a single ear of a person at a single frequency, i.e.,
%%%%%%%%%%%%%%%%%%%%%%%%%%%%%%%%%%%%%%%%%%%%%%%%%%%%%%%%%%%%%%%%%%%%%%%%%%%%
\begin{IEEEeqnarray}{rCl}
\label{eq:big_four}
\hat{P}_{\mathrm{PI}}(\omega,x,y,z)
&=&{\hat{P}^{R}_{\mathrm{PI}}(\omega,x,y>0,z)}         \nonumber\\
&&\cup \; i\cdot{\hat{P}^{I}_{\mathrm{PI}}(\omega,x,y>0,z)}   \nonumber\\
&&\cup \; {\hat{P}^{R}_{\mathrm{PI}}(\omega,x,y<0,z)}   \nonumber\\
&&\cup \; i\cdot{\hat{P}^{I}_{\mathrm{PI}}(\omega,x,y<0,z)},
\end{IEEEeqnarray}
where $\cup$ is the union operator.
%%%%%%%%%%%%%%%%%%%%%%%%%%%%%%%%%%%%%%%%%%%%%%%%%%%%%%%%%%%%%%%%%%%%%%%%%%%%
Denote $(\theta_e,\phi_e)$ as an arbitrary direction, the PINN method estimates 
HRTF for that direction as $\hat{P}_{\mathrm{PI}}(\omega,x_e,y_e,z_e)$,
where $(x_e,y_e,z_e)$ correspond to $(r_\mathrm{h}, \theta_e, \phi_e)$.
%%%%%%%%%%%%%%%%%%%%%%%%%%%%%%%%%%%%%%%%%%%%%%%%%%%%%%%%%%%%%%%%%%%%%%%%%%%%

%%%%%%%%%%%%%%%%%%%%%%%%%%%%%%%%%%%%%%%%%%%%%%%%%%%%%%%%%%%%%%%%%%%%%%%%%%%%%%%%%

%%%%%%%%%%%%%%%%%%%%%%%%%%%%%%%%%%%%%%%%%%%%%%%%%%%%%%%%%%%%%%%%%%%%%%%%%%%%
\subsection{Configuration rationals\label{sec:conf}}
This section provides  the configuration rationals for the proposed PINN method.

\textbf{Cartesian coordinates vs spherical coordinates:}
For the PINN method, HRTF is expressed in Cartesian coordinates instead of spherical 
coordinates for two reasons.
First, the Laplacian in spherical coordinates, Eq.~\eqref{eq:wave1},
can be numerically unstable due to the $\sin\theta$ term in denominators. 
Second, HRTF is evaluated on a single sphere, and thus there is no variations
along the radial direction. 
This makes the PINN unable to estimate the first-order and second-order 
radial gradient used for calculating the Laplacian in spherical coordinates, Eq.~\eqref{eq:wave1}.

\textbf{Frequency-wise upsampling:} 
The proposed method focuses on upsampling HRTF for each frequency.
This allows better control of the training process with respect to frequency.

To model HRTF of a person at all frequencies and at two ears, 
$L_\omega$ $\times$ 2 $\times$ $4= 8\;L_\omega$ PINN methods need to be built,
where $L_\omega$ is the number of frequencies of interest.

%%%%%%%%%%%%%%%%%%%%%%%%%%%%%%%%%%%%%%%%%%%%%%%%%%%%%%%%%%%%%%%%%%%%%%%%%%%%
\textbf{Loss:} 
The data loss $\mathfrak{L}_{\mathrm{data}}(\omega)$ prompts 
the PINN method output to approximate the measured HRTF, i.e.,
$\hat{P}_{\mathrm{PI}}(\omega,x_q,y_q,z_q)\approx{}P(\omega,x_q,y_q,z_q)$ for $q\in[1,Q]$. 
%%%%%%%%%%%%%%%%%%%%%%%%%%%%%%%%%%%%%%%%%%%%%%%%%%%%%%%%%%%%%%%%%%%%%%%%%%%%

%%%%%%%%%%%%%%%%%%%%%%%%%%%%%%%%%%%%%%%%%%%%%%%%%%%%%%%%%%%%%%%%%%%%%%%%%%%
The PDE loss $\mathfrak{L}_{\mathrm{PDE}}(\omega)$ regularizes the PINN 
method output to conform with the Helmholtz equation at $\{(x_d,y_d,z_d)\}_{d=1}^{D}$, 
a super set of $\{(x_q,y_q,z_q)\}_{q=1}^{Q}$.
%%%%%%%%%%%%%%%%%%%%%%%%%%%%%%%%%%%%%%%%%%%%%%%%%%%%%%%%%%%%%%%%%%%%%%%%%%%
This helps the PINN method to generate physically valid output 
beyond the training data. The regularization in Eq.~\eqref{eq:regulation},
on the other hand, may not enable the SH method to generate physically 
valid output as shown in the experiment section. 
%%%%%%%%%%%%%%%%%%%%%%%%%%%%%%%%%%%%%%%%%%%%%%%%%%%%%%%%%%%%%%%%%%%%%%%%%%%

\textbf{Helmholtz equation:} 
Generally speaking,  PINN methods are trying to solve 
multiple loss optimization problems~\cite{multiple_target}. 
Additional parameters, such as $\lambda$ in Eq.~\eqref{eq:conventionalloss}, 
are normally used to balance different loss terms~\cite{multiple_target}. 
%%%%%%%%%%%%%%%%%%%%%%%%%%%%%%%%%%%%%%%%%%%%%%%%%%%%%%%%%%%%%%%%%%%%%%%%%%%%
Although tuning additional parameters may improve the performance of  
PINN methods, we decided not to do so. 

%%%%%%%%%%%%%%%%%%%%%%%%%%%%%%%%%%%%%%%%%%%%%%%%%%%%%%%%%%%%%%%%%%%%%%%%%%%%
Instead, we re-arrang the Helmholtz equation Eq. \eqref{eq:wave0} to be  
\begin{IEEEeqnarray}{rcl}
\label{eq:wc_helmholtz}
\frac{\nabla^2P}{(\omega/c)^2} + P=0.    
\end{IEEEeqnarray}
%where the $(\omega/c)^2$ term is used as the 
%adenominator for the Laplacian $\nabla^2P$ rather 
%than as a multiplier for the sound pressure $P$. 
Corresponding modifications are made to the PDE loss in Eq.~\eqref{eq:cost}. 
Under the modification, the PDE loss will have the same physical unit 
as the data loss, and hence balancing is not needed. 
%%%%%%%%%%%%%%%%%%%%%%%%%%%%%%%%%%%%%%%%%%%%%%%%%%%%%%%%%%%%%%%%%%%%%%%%%%%%
Without the need for tuning loss-balancing parameters~\cite{multiple_target}, 
the training of the PINN method is simplified.
%%%%%%%%%%%%%%%%%%%%%%%%%%%%%%%%%%%%%%%%%%%%%%%%%%%%%%%%%%%%%%%%%%%%%%%%%%%%

%%%%%%%%%%%%%%%%%%%%%%%%%%%%%%%%%%%%%%%%%%%%%%%%%%%%%%%%%%%%%%%%%%%%%%%%%%%%
\textbf{PINN method width:}
Based on our knowledge of SHs, we provide guidance on the PINN method width,
the number of neurons in each hidden layer.

%%%%%%%%%%%%%%%%%%%%%%%%%%%%%%%%%%%%%%%%%%%%%%%%%%%%%%%%%%%%%%%%%%%%%%%%%%%%%%%%
Although HRTF is defined over all directions, 
the most interesting directions are on the $xy$ plane~\cite{hrtf_mea}, 
where the SH decomposition of HRTF reduces to 
\begin{IEEEeqnarray}{rCl}
\label{eq:circle}
P(\omega,\theta=0,\phi)
&\approx&
\sum_{u=0}^{U}\sum_{v=-u}^{u}A_{u,v}(\omega)Y_{u}^{v}(0,\phi)         \nonumber\\
&=&
\sum_{u=0}^{U}\sum_{v=-u}^{u}A_{u,v}(\omega)                        \nonumber\\
&&\times 
\sqrt{\frac{(2\mu+1)(u-|v|)!}{4\pi(u+|v|)!}}
\mathcal{P}_{u}^{|v|}(0)e^{iv\phi}                          \nonumber\\
&=&
\sum_{v=-U}^{U}
A_{v}(\omega)
%A_{u,v}(\omega)                        \nonumber\\
%&&\times 
%\sqrt{\frac{(2\mu+1)(\mu-|\nu|)!}{4\pi(\mu+|\nu|)!}}
%\mathcal{P}_{u}^{v}(0)
e^{iv\phi},  %                        \nonumber\\
\end{IEEEeqnarray}  
and
\begin{IEEEeqnarray}{rcl}
A_{v}(\omega)&=&
\sum_{u=|v|}^{U}A_{u,v}(\omega)   
\sqrt{\frac{(2u+1)(u-|v|)!}{4\pi(u+|v|)!}}
\mathcal{P}_{u}^{|v|}(0),\qquad
\end{IEEEeqnarray}  
is obtained by manipulating the second and third lines of Eq.~\eqref{eq:circle}.
Eq.~\eqref{eq:circle} indicates that HRTF on the $xy$ plane can be expressed 
by $2U+1$  basis functions with related weights 
$\{e^{iv\phi}, A_v(\omega), v\in[-U, U]\}$.

Based on Fig.~\ref{fig:pinn}, 
%if the PINN learns the distribution of the underlying HRTFs, 
from the output's point of view, HRTF on the $xy$ plane can be expressed as 
%by the output of 
%$W$ neurons together with corresponding weights
\begin{IEEEeqnarray}{rCl}
\label{eq:activate}
\hat{P}_\mathrm{PI}(\omega, x, y ,z=0) 
&=& \sum_{j=1}^{W} \sigma_j \mathrm{w}_j    
%\nonumber\\
%&&
+  
\sigma_0b, 
\end{IEEEeqnarray}
where $\sigma_j$ denotes the value of the $j$-th neuron on the last hidden 
layer, $\mathrm{w}_j$ is the corresponding weight, 
and the bias $b$ can be regarded as the weight for a linear activation function $\sigma_0$. 
$\{\sigma_j\}_{j=0}^W$ are implicit functions of Cartesian coordinates.

If the PINN method learns the underlying HRTF, then for corresponding 
$\phi$ and $(x, y)$ there must be 
\begin{IEEEeqnarray}{rCl}
\label{eq:equal}
P(\omega,\theta=0,\phi)&\approx&\hat{P}_\mathrm{PI}(\omega, x, y ,z=0),
\end{IEEEeqnarray}
and hence  
\begin{IEEEeqnarray}{rCl}
\label{eq:basis_equal}
\sum_{v=-U}^{U}
A_{v}(\omega)
e^{iv\phi}
&\approx& \sum_{j=1}^{W} \sigma_j \mathrm{w}_j    
+  
\sigma_0b, 
\end{IEEEeqnarray}
though they are expressed in different coordinates and with different 
decompositions.   
Based on Eq.~\eqref{eq:basis_equal} and the minimum description length principle~\cite{minimum}, 
we choose the PINN method width as 
\begin{IEEEeqnarray}{rcl}
W+1= 2U+1, \quad \mathrm{or}\quad  W = 2U.   
\end{IEEEeqnarray}
%%%%%%%%%%%%%%%%%%%%%%%%%%%%%%%%%%%%%%%%%%%%%%%%%%%%%%%%%%%%%%%%%%%%%%%%%%%%%%%%

In Sec.~\ref{sec:proposed}, four identical PINN methods are used for 
modeling the left/right and real/imaginary parts of HRTF. 
This prompts us to arrive at the final choice for the width 
(the number of neurons in each hidden layer) of a PINN method as 
\begin{IEEEeqnarray}{rCl}
\label{eq:width}
W&=&2U/4                    \nonumber\\                       
&=&  U/2                         \nonumber\\
%\approx 
%\begin{cases}
%\lceil{f/500}\rceil, & f\leq3 \mathrm{kHz},\\ 
%\lceil{f/1000}\rceil, & f>3 \mathrm{kHz}. 
%\end{cases}
&\approx&
\begin{cases}
\lceil{f/500}\rceil, & f<3\; \mathrm{kHz},         \\ 
6, & 3 \;\mathrm{kHz} \leq f\leq6 \;\mathrm{kHz},       \\ 
\lceil{f/1000}\rceil, & f > 6 \;\mathrm{kHz}. 
\end{cases} 
\end{IEEEeqnarray}
%%%%%%%%%%%%%%%%%%%%%%%%%%%%%%%%%%%%%%%%%%%%%%%%%%%%%%%%%%%%%%%%%%%%
The width $W$ as a function of frequency is presented in Fig.~\ref{fig:dimensionality} 
for reference.
%,  where we let $W=\max\{  \lceil{}f/500\rceil, \lceil{}f /1000\rceil\}$ 
%for 3 kHz $<f<$ 6 kHz \cite{hrtf_insight}. 
%%%%%%%%%%%%%%%%%%%%%%%%%%%%%%%%%%%%%%%%%%%%%%%%%%%%%%%%%%%%%%%%%%%%
Note that similar to Eqs. \eqref{eq:order} and \eqref{eq:size},
Eq.~\eqref{eq:width} represents approximations only.
%%%%%%%%%%%%%%%%%%%%%%%%%%%%%%%%%%%%%%%%%%%%%%%%%%%%%%%%%%%%%%%%%%%%
%Indeed, for HRTF upsampling, we are facing a data insufficient 
%condition, and to avoid the over-fitting problem we may need to 
%slightly reduce the width $W$.  
%%%%%%%%%%%%%%%%%%%%%%%%%%%%%%%%%%%%%%%%%%%%%%%%%%%%%%%%%%%%%%%%%%%%

\textbf{PINN depth:}
%%%%%%%%%%%%%%%%%%%%%%%%%%%%%%%%%%%%%%%%%%%%%%%%%%%%%%%%%%%%%%%%%%%%
For the proposed PINN method, with no additional parameters for different 
loss terms in Eq.~\eqref{eq:cost}, 
the activation function fixed to be $\tanh$, and the width 
set to be $W=U/2$, the last parameter that could determine
its performance is the depth $L$ (the number of hidden layers), 
other than the training data.
%%%%%%%%%%%%%%%%%%%%%%%%%%%%%%%%%%%%%%%%%%%%%%%%%%%%%%%%%%%%%%%%%%%%
We found that a depth of $L=3$ is a suitable choice, which balances 
the upsampling accuracy with the model complexity. 
%as shown in the 
%experiment and discussion sections. 
This may be because the Helmholtz equation is a second-order PDE or 
three variables $(x,y,z)$ are needed to determine the HRTF for a direction.
%%%%%%%%%%%%%%%%%%%%%%%%%%%%%%%%%%%%%%%%%%%%%%%%%%%%%%%%%%%%%%%%%%%%
In future studies, we plan to further analyze the Helmholtz equation 
and its solutions to determine an optimal depth $L$. 
%%%%%%%%%%%%%%%%%%%%%%%%%%%%%%%%%%%%%%%%%%%%%%%%%%%%%%%%%%%%%%%%%%%%

%%%%%%%%%%%%%%%%%%%%%%%%%%%%%%%%%%%%%%%%%%%%%%%%%%%%%%%%%%%%%%%%%%%%%%%%%%%%
\begin{figure}[t]
\centering
\includegraphics[width=9.cm]{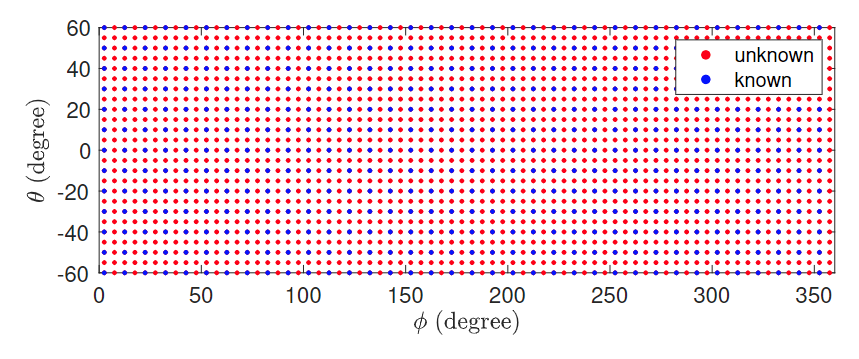}
\caption{Directions of the known HRTF and the unknown HRTF for the interpolation experiment.}
\label{fig:int_sampling}
\end{figure}

%%%%%%%%%%%%%%%%%%%%%%%%%%%%%%%%%%%%%%%%%%%%%%%%%%%%%%%%%%%%%%%%%%%%%%%%%%%%
\section{\label{sec:int_exp}Interpolation experiment}
Numerical experiments were conducted in this section to interpolate 
unknown HRTF whose direction is between those of measured ones. 

\subsection{Data processing}
%%%%%%%%%%%%%%%%%%%%%%%%%%%%%%%%%%%%%%%%%%%%%%%%%%%%%%%%%%%%%%%%%%%%%%%%%%
Experiments were conducted on the HUTUBS dataset~\cite{hutubs}, 
subject 11, 12, ... 50, left-ear HRTF. 
Based on SH coefficients up to 35-th order~\cite{hutubs}, HRTF for 
1260 directions, where $\theta\in[-60^{\circ}, -54^{\circ}, ..., 
60^{\circ}]$ and $\phi\in[4^{\circ}, 10^{\circ}, ..., 358^{\circ}]$,
was calculated according to Eq.~(4).  
One forth of the HRTF was set as the known HRTF, and the rest was 
set as the unknown HRTF (ground truth). 
Directions of the known HRTF and the unknown HRTF were shown in Fig.~4.
%%%%%%%%%%%%%%%%%%%%%%%%%%%%%%%%%%%%%%%%%%%%%%%%%%%%%%%%%%%%%%%%%%%%
The magnitudes of all HRTFs were normalized to be within $[0, 1]$. 

%With 330 known HRTFs, the SH method could at most determine the SH coefficients 
%up to order $U=\lfloor{\sqrt{330/2}}\rfloor-1=17$~\cite{Rafaely_book}
%($\lfloor\cdot\rfloor$ is the floor operation). 
%or corresponding  up to 
%frequency $f_\mathrm{max}={Uc}/(2\pi{r_h)}\approx$ 10 kHz. 

The 256 tap head-related impulse response~\cite{hutubs}, sampled at 44100 Hz, 
was transformed into frequency domain through discrete Fourier transform, 
resulting corresponding HRTF.
HRTF was evaluated at [2067, 4134, 6202, 8269, 10336, 12403, 14470] Hz, which 
approximate multiples of the base frequency $44100/256$ Hz.
Hereafter, we referred to these frequencies as 2.1, 4.1, 6.2, 8.2, 10.3, 12.3, 
and 14.4 kHz for notational simplicity. 
%%%%%%%%%%%%%%%%%%%%%%%%%%%%%%%%%%%%%%%%%%%%%%%%%%%%%%%%%%%%%%%%%%%%%%%%%%
HRTF was not evaluated in higher frequencies ($f>14.4$ kHz) because they do not
contribute significantly to the perception of source location~\cite{hrtf_mea}. 
We have evaluated the interpolation performance of different methods for $f<2.1$ kHz. Similar to Fig.~6 (a), in such low frequency range the interpolation error of the SH method is less than those of the PINN method, the NN method, and the HRTF field method. 
The results were not shown for brevity. 
%%%%%%%%%%%%%%%%%%%%%%%%%%%%%%%%%%%%%%%%%%%%%%%%%%%%%%%%%%%%%%%%%%%%%%%%%%

\subsection{Implementation}
%%%%%%%%%%%%%%%%%%%%%%%%%%%%%%%%%%%%%%%%%%%%%%%%%%%%%%%%%%%%%%%%%%%%%%%%%%%%
\textbf{SH method:} 
We implemented the SH method~\cite{SH2004} following Eqs.~\eqref{eq:SH_first} 
- \eqref{eq:sh_last}. 
%%%%%%%%%%%%%%%%%%%%%%%%%%%%%%%%%%%%%%%%%%%%%%%%%%%%%%%%%%%%%%%%%%%%%%%%%%%%

%%%%%%%%%%%%%%%%%%%%%%%%%%%%%%%%%%%%%%%%%%%%%%%%%%%%%%%%%%%%%%%%%%%%%%%%%%%%
\textbf{PINN method:} 
%%%%%%%%%%%%%%%%%%%%%%%%%%%%%%%%%%%%%%%%%%%%%%%%%%%%%%%%%%%%%%%%%%%%%%%%%%%%
%Let $(\theta,\phi)$ be the direction of HRTFs. 
%The spherical coordinates $(r_\mathrm{h},\theta,\phi)$ are transformed 
%into Cartesian coordinates. 
The known HRTF and corresponding Cartesian coordinates were the training 
data pairs for calculating the data loss $\mathfrak{L}_{\mathrm{Data}}(\omega)$. 
Cartesian coordinates of the known HRTF and the unknown HRTF were combined 
and used for calculating the PDE loss $\mathfrak{L}_\mathrm{PDE}(\omega)$.
%%%%%%%%%%%%%%%%%%%%%%%%%%%%%%%%%%%%%%%%%%%%%%%%%%%%%%%%%%%%%%%%%%%%%%%%%%

%%%%%%%%%%%%%%%%%%%%%%%%%%%%%%%%%%%%%%%%%%%%%%%%%%%%%%%%%%%%%%%%%%%%%%%%%%
%Four PINNs were used for modeling the HRTFs for one ear of a person at one 
%frequency. The outputs of four PINNs were combined according to Eq.~\eqref{eq:big_four} 
%to arrive at the complex value HRTF estimation $\hat{P}_{\mathrm{PI}}(\omega, x, y, z)$. 
%%%%%%%%%%%%%%%%%%%%%%%%%%%%%%%%%%%%%%%%%%%%%%%%%%%%%%%%%%%%%%%%%%%%%%%%%%

%%%%%%%%%%%%%%%%%%%%%%%%%%%%%%%%%%%%%%%%%%%%%%%%%%%%%%%%%%%%%%%%%%%%%%%%%%
To investigate how depth $L$ and width $W$ influence the interpolation performance, 
we implemented the PINN method in seven cases: 
\begin{enumerate}
\item 
$L=2$, $W=U/2$;   
\item 
$L=2$, $W=U$;  
\item 
$L=3$, $W=U/2$;  
\item 
$L=3$, $W=U$;  
\item 
$L=4$, $W=U/2$;  
\item 
$L=4$, $W=U$;  
\item 
$L=4$, $W=50$.  
\end{enumerate}
For the first six cases, the width $W$ varied with frequencies. 
For the last case, the width $W$ was fixed. 
%%%%%%%%%%%%%%%%%%%%%%%%%%%%%%%%%%%%%%%%%%%%%%%%%%%%%%%%%%%%%%%%%%%%%%%%%%

%%%%%%%%%%%%%%%%%%%%%%%%%%%%%%%%%%%%%%%%%%%%%%%%%%%%%%%%%%%%%%%%%%%%%%%%%%
We used the Tensorflow library for training, initialized the trainable 
parameters according to the Xavier initialization~\cite{init}, set the 
activation function to be $\tanh$, chose the ADAM optimizer with a learning 
rate of 0.001, and trained the PINN methods for $10^6$ epochs.
%%%%%%%%%%%%%%%%%%%%%%%%%%%%%%%%%%%%%%%%%%%%%%%%%%%%%%%%%%%%%%%%%%%%%%%%%%
%Case 1), 2), 3), and 4), were trained on one CPU core of a laptop with M1 
%Pro chip. Case 5) was trained on an Nvidia 2080Ti GPU.
%%%%%%%%%%%%%%%%%%%%%%%%%%%%%%%%%%%%%%%%%%%%%%%%%%%%%%%%%%%%%%%%%%%%%%%%%%
%For all cases, the training for one person's HRTFs at seven frequencies
%took about 5 to 7 hours. 
%%%%%%%%%%%%%%%%%%%%%%%%%%%%%%%%%%%%%%%%%%%%%%%%%%%%%%%%%%%%%%%%%%%%%%%%%%

\textbf{NN method:} 
%%%%%%%%%%%%%%%%%%%%%%%%%%%%%%%%%%%%%%%%%%%%%%%%%%%%%%%%%%%%%%%%%%%%%%%%%%
Another method was implemented identical to the PINN method, except that the 
PDE loss $\mathfrak{L}_\mathrm{PDE}(\omega)$ was not calculated and thus 
was not used for regularizing the network output. Hereafter, this method 
was denoted as the NN method. The NN method was implemented with different 
numbers of hidden layers and neurons same to the first four cases of the 
PINN method.
%%%%%%%%%%%%%%%%%%%%%%%%%%%%%%%%%%%%%%%%%%%%%%%%%%%%%%%%%%%%%%%%%%%%%%%%%%
%?gThe training procedure involved joint-optimization of the latent code $\mathbf{z}$ 
%?gand network's parameters. 
%?gInitially, the latent code $\mathbf{z}$ was initialized as zeros and inferred by computing the gradient of the Mean Square Error (MSE) between the HRTF estimation and the ground-truth for the subject across all directions $\mathbf{x}$. 
%Subsequently, 

\textbf{HRTF field method:} 
%%%%%%%%%%%%%%%%%%%%%%%%%%%%%%%%%%%%%%%%%%%%%%%%%%%%%%%%%%%%%%%%%%%%%%%%%%
An additional HRTF field method~\cite{neuralfield} was implemented using the Sinusoidal 
Representation Network (SIREN) architecture, which is also a multi-layer fully connected 
feed-forward neural network but using $\sine$ as the activation function. 
%%%%%%%%%%%%%%%%%%%%%%%%%%%%%%%%%%%%%%%%%%%%%%%%%%%%%%%%%%%%%%%%%%%%%%%%%%
%A latent code $\mathbf{z}$ with a dimension of 32 served as the intermediary between 
%an individual's head geometry and corresponding HRTF. 
%%%%%%%%%%%%%%%%%%%%%%%%%%%%%%%%%%%%%%%%%%%%%%%%%%%%%%%%%%%%%%%%%%%%%%%%%%%%
%The method was denoted as $G: \mathbb{R}^{2+32} \mapsto \mathbb{R}^K$, which 
%mapped direction $(\theta, \phi)$ and the latent code $\mathbf{z}$ to an individual's HRTF 
%$P(\omega,\theta, \phi)$.
%%%%%%%%%%%%%%%%%%%%%%%%%%%%%%%%%%%%%%%%%%%%%%%%%%%%%%%%%%%%%%%%%%%%%%%%%%
We found that the method was unable to converge if trained with less than 70 subjects' HRTF data, and thus trained the method for three times. In the first time, left-ear HRTF of subject 1-10 and 25-96 was used as 
training set, and left-ear HRTF of subject 11-24 was used as testing set. 
In the second time, left-ear HRTF of subject 1-24 and 40-96 was 
used as training set, and left-ear HRTF of subject 25-39 was used as testing set. 
In the third time, left-ear HRTF of subject 1-35 and 51-96 was 
used as training set, and left-ear HRTF of subject 36-50 was used as testing set. 
%%%%%%%%%%%%%%%%%%%%%%%%%%%%%%%%%%%%%%%%%%%%%%%%%%%%%%%%%%%%%%%%%%%%%%%%%%
The three time training was conducted because we found that the method was 
unable to converge if trained with less than 70 subjects' HRTF data. 
%%%%%%%%%%%%%%%%%%%%%%%%%%%%%%%%%%%%%%%%%%%%%%%%%%%%%%%%%%%%%%%%%%%%%%%%%%
For each time, 
the training data consisted of 330 known HRTFs and corresponding 
Cartesian coordinates from the test set, along with 1260 known plus unknown HRTFs 
and corresponding Cartesian coordinates from the training set; 
%%%%%%%%%%%%%%%%%%%%%%%%%%%%%%%%%%%%%%%%%%%%%%%%%%%%%%%%%%%%%%%%%%%%%%%%%%
the testing data consisted of 1260 known and unknown HRTFs and corresponding Cartesian 
coordinates from the test set. 
%%%%%%%%%%%%%%%%%%%%%%%%%%%%%%%%%%%%%%%%%%%%%%%%%%%%%%%%%%%%
We set the learning rate following~\cite{neuralfield} and trained for 240 epochs. 
%monitoring loss reduction during training. 
Please refer to \cite{neuralfield} for the theory and implementation of the 
HRTF field method. 
%%%%%%%%%%%%%%%%%%%%%%%%%%%%%%%%%%%%%%%%%%%%%%%%%%%%%%%%%%%%

\subsection{Performance evaluation}
%%%%%%%%%%%%%%%%%%%%%%%%%%%%%%%%%%%%%%%%%%%%%%%%%%%%%%%%%%%%%%%%%%%%%%%%%%%%
The performance of each method was evaluated based on the interpolation error
\begin{IEEEeqnarray}{rcl}
\mathcal{E}(\omega)&=& 20\log_{10} 
\frac{\sum_{e=1}^{930} | P(\omega,\theta_e,\phi_e) - \hat{P}(\omega,\theta_e,\phi_e)
|}{\sum_{e=1}^{930}|P(\omega,\theta_e,\phi_e)|}, \quad
\end{IEEEeqnarray}
where 
$P(\omega,\theta_e,\phi_e)$ and $\hat{P}(\omega,\theta_e,\phi_e)$ 
were the unknown HRTF (ground-truth) and its estimation generated by different 
methods at directions $\{(\theta_e,\phi_e)\}_{e=1}^{930}$, respectively. 
%%%%%%%%%%%%%%%%%%%%%%%%%%%%%%%%%%%%%%%%%%%%%%%%%%%%%%%%%%%%%%%%%%%%%%%%%%%%

\subsection{Result: at a frequency for one subject\label{sec:int_visual}}
\begin{figure}[]
\centering
\includegraphics[width=9.cm]{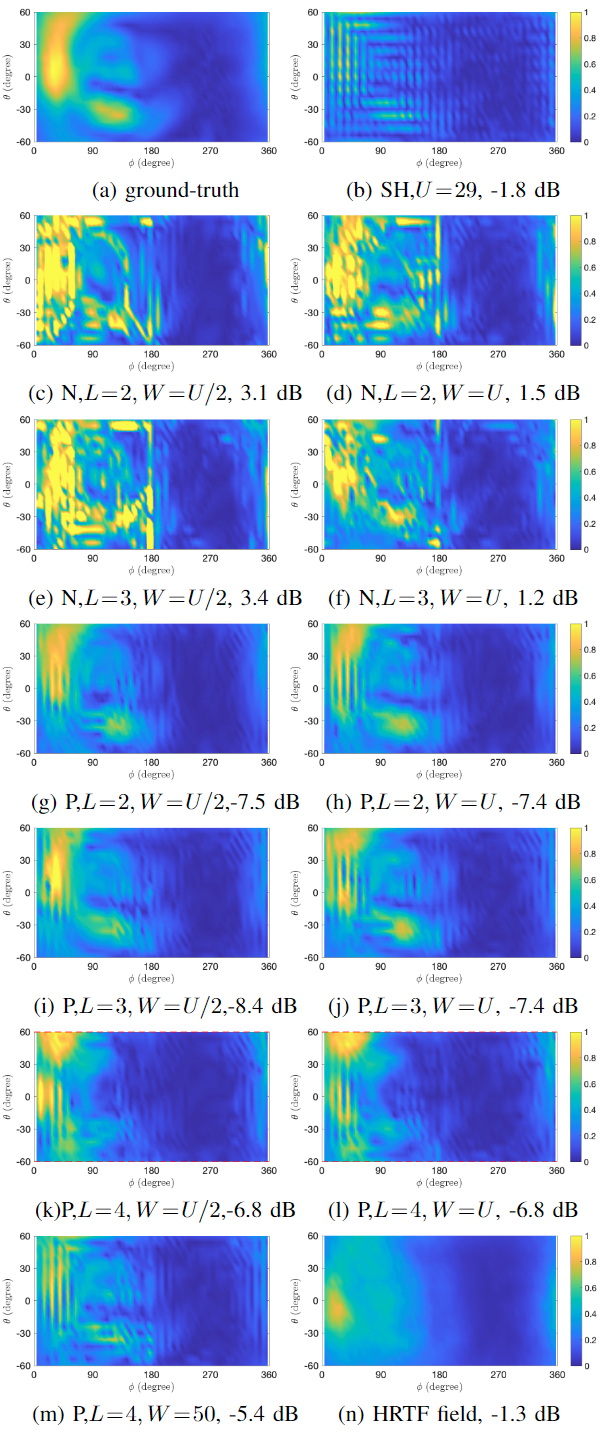}
\caption{Interpolation results: left-ear HRTF of subject 40 at 14.4 kHz,
magnitudes of the ground-truth and interpolations by different methods.}
%ground-truth, interpolations and interpolation errors of different methods.}
\label{fig:int_sphere}
\end{figure}

We presented an interpolation result in Fig.~\ref{fig:int_sphere},
which showed magnitudes of the left-ear HRTF of subject 
40 at 14.4 kHz, ground-truth, interpolations and interpolation errors 
of different methods. 

Note that, in captions of Figs.~\ref{fig:int_sphere}, \ref{fig:int_array}, \ref{fig:ext_sphere}, 
and \ref{fig:ext_array}, we used ``P, $L=3, W=U/2$, -8.4 dB" to 
denote that the PINN method with depth $L=3$ and width $W=U/2$ achieved 
interpolation error of -8.4 dB. Other captions could be interpreted similarly.

In the case, the dimensionality of HRTF under SH decomposition was 
$U=\lceil{2\pi{}f{}r_\mathrm{h}}/c\rceil{}=29$. 
%and thus HRTFs for at least 
%(U+1)2=960(U+1)^2=960 directions, are needed to determine the SH coefficients. 
With only $330\ll(29+1)^2$ known HRTFs, the SH method was unable 
determine the SH coefficients up to $U=30$, and hence was unable to 
accurately interpolate the unknown HRTF.
%Comparing Figure~\ref{fig:int_sphere} (b) to Fig.~\ref{fig:int_sampling},  we saw that the SH method 

%The NN methods, which were solely trained on the known HRTFs, were unable to 
%interpolate the unknown HRTFs. 

Referring to Fig.~\ref{fig:int_sampling} and Fig.~\ref{fig:int_sphere} (c) - (f), 
we saw the NN methods assigned physically invalid values to the unknown HRTFs, 
and the interpolation errors were larger than $0$ dB in all four cases.

Thanks to the regularization of the PDE loss, the PINN methods did not assigned 
physically invalid values to the unknown HRTFs as shown in Fig.~\ref{fig:int_sphere} (g) - (m). 
Nonetheless, the PINN methods tended to assign zero to the unknown HRTFs, 
especially the deeper and wider PINN method shown in Fig.~\ref{fig:int_sphere} (m). 
This indicated over-fitting and demonstrated the difficult of training a PINN~\cite{pinn_pp1,pinn_pp2,pinn_pp3}.
Nonetheless, the PINN method with depth $L=3$ and width $W=U/2$ showed 
the least interpolation error of -8.4 dB. 

As shown in Fig.~\ref{fig:int_sphere} (n),
the HRTF field method failed to capture the structure of HRTF, 
and the interpolation error was about $-1.3$ dB. 

\subsection{Result: across all frequencies and subjects\label{sec:int_array}}
\begin{figure}[!ht]
\centering
\includegraphics[width=8.6cm]{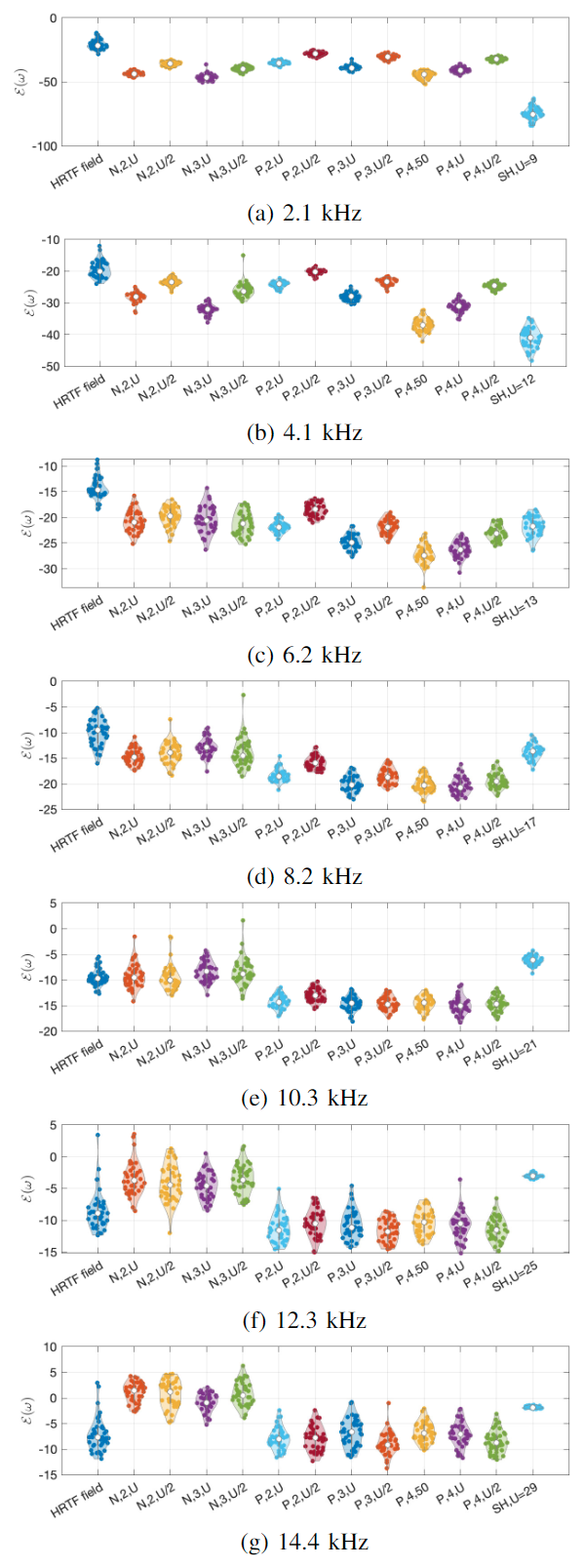}
\caption{Interpolation errors across all frequencies and subjects.
}
\label{fig:int_array}
\end{figure}

Figure~\ref{fig:int_array} showed the interpolation errors across all 
frequencies and subjects. 
%As shown in Fig.~\ref{fig:int_array}, the 
%interpolation errors of all methods deteriorate with the increment of 
%frequency.

The interpolation errors of the SH method were the lowest among all methods
in lower frequency range, where $f=2.1, 4.1$ kHz. With the increment of
frequency, the SH method interpolation error increased as well as the interpolation 
errors all other methods.

The NN method interpolation errors of four cases were similar across all
frequencies. Unlike other methods, the interpolation errors of NN methods 
could go above $0$ dB. 

\clearpage

%%%%%%%%%%%%%%%%%%%%%%%%%%%%%%%%%%%%%%%%%%%%%%%%%%%%%%%%%%%%

%%%%%%%%%%%%%%%%%%%%%%%%%%%%%%%%%%%%%%%%%%%%%%%%%%%%%%%%%%%%%%%%%%%%%%%%%%%%%%%%%%%%%%%

%%%%%%%%%%%%%%%%%%%%%%%%%%%%%%%%%%%%%%%%%%%%%%%%%%%%%%%%%%%%%%%%%%%%%%%%%%
For frequency $f\leq 4.1$ kHz, the PINN method interpolation errors were 
similar to those of the NN methods. 
%%%%%%%%%%%%%%%%%%%%%%%%%%%%%%%%%%%%%%%%%%%%%%%%%%%%%%%%%%%%%%%%%%%%%%%%%%
However, for $f\geq 8.2$ kHz, the PINN method interpolation errors 
were lower than those of the NN methods. 
%%%%%%%%%%%%%%%%%%%%%%%%%%%%%%%%%%%%%%%%%%%%%%%%%%%%%%%%%%%%%%%%%%%%%%%%%%
In high-frequency range where $f\geq10.3$ kHz, the PINN method with $L=3$ 
and $W=U/2$ demonstrated the smallest average interpolation errors among the 
40 subjects for about $-14$ dB at 10.3 kHz, about $-12.5$ dB at 12.3 kHz, 
and about $-9$ dB at $-14.4$ kHz. 
%%%%%%%%%%%%%%%%%%%%%%%%%%%%%%%%%%%%%%%%%%%%%%%%%%%%%%%%%%%%%%%%%%%%%%%%%%

The interpolation errors of the HRTF field method showed the least variance 
across the frequencies. However, it did not achieved the smallest 
interpolation errors at any frequency.

\section{Extrapolation experiment\label{sec:ext_exp}}
Numerical experiments were conducted in this section to extrapolate 
the unknown HRTF whose direction is beyond those of the measured ones.

\subsection{Data processing}
\begin{figure}[t]
\centering
\includegraphics[width=9.0cm]{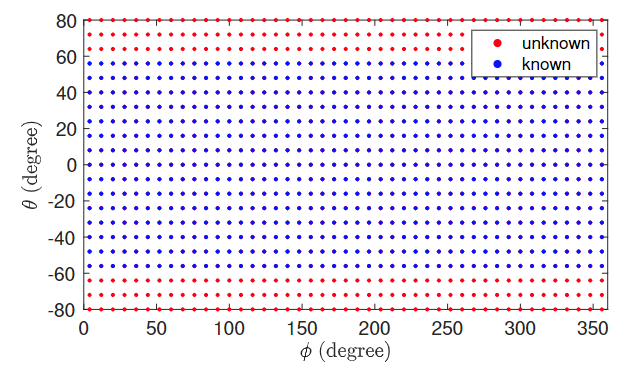}
\caption{Directions of the known HRTF and the unknown HRTF for the extrapolation experiment.}
\label{fig:ext_sampling}
\end{figure}

%%%%%%%%%%%%%%%%%%%%%%%%%%%%%%%%%%%%%%%%%%%%%%%%%%%%%%%%%%%%%%%%%%%%%%%%%%
Experiments were also conducted on the HUTUBS dataset~\cite{hutubs}, 
subject 11 to 50, left-ear HRTF. 
Based on SH coefficients up to 35-th order~\cite{hutubs}, 
HRTF for 675 directions,
where $\theta\in[-56^{\circ}, -48^{\circ}, ..., 56^{\circ}]$ and 
$\phi\in[4^{\circ}, 12^{\circ}, ..., 356^{\circ}]$, 
were calculated according to Eq.~\eqref{eq:SH_first} and used as the known HRTF;
%%%%%%%%%%%%%%%%%%%%%%%%%%%%%%%%%%%%%%%%%%%%%%%%%%%%%%%%%%%%%%%%%%%%%%%%%%
HRTF for 270 directions, where $\theta\in[-80^{\circ}, -72^{\circ}, -64^{\circ}, 
64^\circ, 72^\circ, 80^\circ]$ and $\phi\in[4^{\circ}, 12^{\circ}, ..., 356^{\circ}]$, 
were calculated according to Eq.~\eqref{eq:SH_first} and used as the unknown 
HRTF (ground-truth).
%%%%%%%%%%%%%%%%%%%%%%%%%%%%%%%%%%%%%%%%%%%%%%%%%%%%%%%%%%%%%%%%%%%%%%%%%%
Directions of the known HRTF and the unknown HRTF for the extrapolation experiment
were shown in Fig.~\ref{fig:ext_sampling}.
%%%%%%%%%%%%%%%%%%%%%%%%%%%%%%%%%%%%%%%%%%%%%%%%%%%%%%%%%%%%%%%%%%%%%%%%%%
%The known HRTFs account for approximately one forth of the unknown 
%HRTFs to be interpolated.
%%%%%%%%%%%%%%%%%%%%%%%%%%%%%%%%%%%%%%%%%%%%%%%%%%%%%%%%%%%%%%%%%%%%%%%%%%
HRTF was evaluated at the same frequencies as in Sec.~\ref{sec:int_exp}. 
%%%%%%%%%%%%%%%%%%%%%%%%%%%%%%%%%%%%%%%%%%%%%%%%%%%%%%%%%%%%%%%%%%%%%%%%%%
The magnitudes of all HRTFs were normalized to be within $[0, 1]$. 
%%%%%%%%%%%%%%%%%%%%%%%%%%%%%%%%%%%%%%%%%%%%%%%%%%%%%%%%%%%%%%%%%%%%%%%%%%

\subsection{Implementation}
%%%%%%%%%%%%%%%%%%%%%%%%%%%%%%%%%%%%%%%%%%%%%%%%%%%%%%%%%%%%%%%%%%%%%%%%%%%%
\textbf{SH method:} 
The SH method~\cite{SH2004} was implemented following Eqs.~\eqref{eq:SH_first} 
- \eqref{eq:sh_last}, and set 
$\gamma=10^{-6}, 10^{-4}, 10^{-1}, 10^{-1}, 10^{-2}, 10^{-1}, 10^{-3}$ 
for $f=$ 2.1, 4.1, 6.2, 8.2, 10.3, 12.3, 14.4 kHz in Eq. \eqref{eq:regulation} 
according to a trial-and-error process. 
%%%%%%%%%%%%%%%%%%%%%%%%%%%%%%%%%%%%%%%%%%%%%%%%%%%%%%%%%%%%%%%%%%%%%%%%%%%%

%%%%%%%%%%%%%%%%%%%%%%%%%%%%%%%%%%%%%%%%%%%%%%%%%%%%%%%%%%%%%%%%%%%%%%%%%%%%
\textbf{PINN method and NN method:} 
%%%%%%%%%%%%%%%%%%%%%%%%%%%%%%%%%%%%%%%%%%%%%%%%%%%%%%%%%%%%%%%%%%%%%%%%%%
The implementations of the PINN method and the NN method 
were identical to Sec.~\ref{sec:int_exp} B, except the training data and output. 

\textbf{HRTF field method:} 
The implementation of the HRTF field method~\cite{neuralfield} was similar 
to Sec.~\ref{sec:int_exp} B, except the training and testing data.
%%%%%%%%%%%%%%%%%%%%%%%%%%%%%%%%%%%%%%%%%%%%%%%%%%%%%%%%%%%%
Training data consisted of the 675 known HRTFs and corresponding Cartesian coordinates 
from the test set, 
along with 945 known plus unknown HRTFs and corresponding Cartesian coordinates 
from the training set. 
%%%%%%%%%%%%%%%%%%%%%%%%%%%%%%%%%%%%%%%%%%%%%%%%%%%%%%%%%%%%
Testing data consisted of 945 known plus unknown HRTFs and 
corresponding Cartesian coordinates from the test set.
%%%%%%%%%%%%%%%%%%%%%%%%%%%%%%%%%%%%%%%%%%%%%%%%%%%%%%%%%%%%
Please refer to \cite{neuralfield} for the theory and implementation of the 
HRTF field method.

%%%%%%%%%%%%%%%%%%%%%%%%%%%%%%%%%%%%%%%%%%%%%%%%%%%%%%%%%%%%%%%%%%%%%%%%%%%%
%%%%%%%%%%%%%%%%%%%%%%%%%%%%%%%%%%%%%%%%%%%%%%%%%%%%%%%%%%%%%%%%%%%%%%%%%%%%

%%%%%%%%%%%%%%%%%%%%%%%%%%%%%%%%%%%%%%%%%%%%%%%%%%%%%%%%%%%%%%%%%%%%%%%%%%%%
\begin{figure}[t]
\centering
\includegraphics[width=9.0cm]{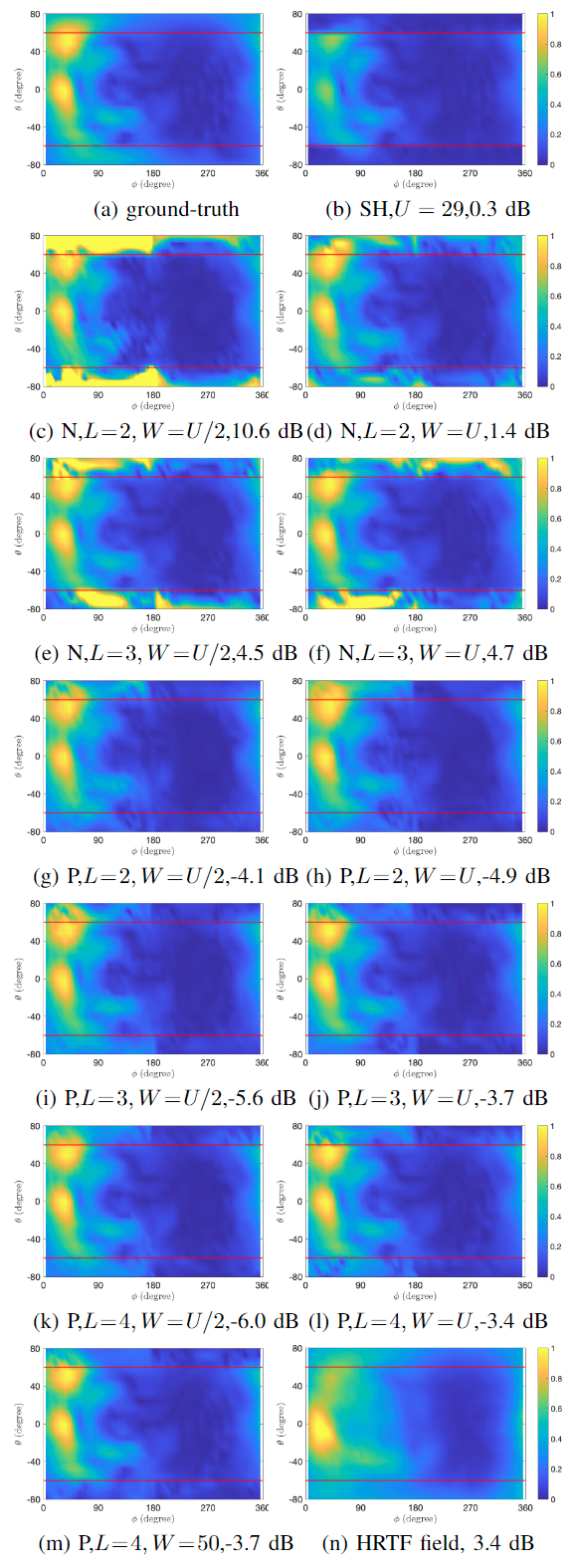}
%%%%%%%%%%%%%%%%%%%%%%%%%%%%%%%%%%%%%%%%%%%%%%%%%%%%%%%%%%%%
\caption{Extrapolation results: left-ear HRTF subject 
20 at 14.4 kHz, magnitudes of the ground-truth and extrapolations by different methods. 
The red lines denote the boundaries between the known HRTF and the unknown HRTF.} 
\label{fig:ext_sphere}
\end{figure}
%%%%%%%%%%%%%%%%%%%%%%%%%%%%%%%%%%%%%%%%%%%%%%%%%%%%%%%%%%%%%%%%%%%%%%%%%%%%%%%%%%%%%%
\begin{figure}[t]
\centering
\includegraphics[width=9.0cm]{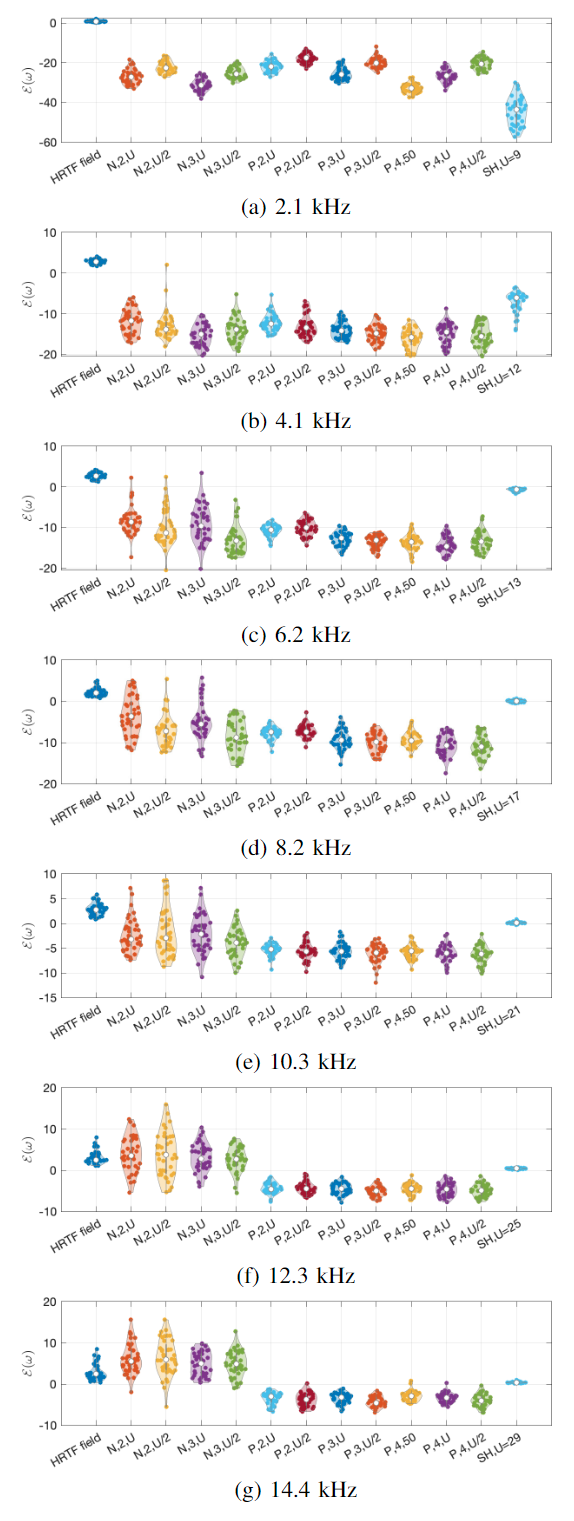}
\caption{Extrapolation result across all frequencies and subjects.}
\label{fig:ext_array}
\end{figure}

\subsection{Performance evaluation}
%%%%%%%%%%%%%%%%%%%%%%%%%%%%%%%%%%%%%%%%%%%%%%%%%%%%%%%%%%%%%%%%%%%%%%%%%%%%
The performance of each method was evaluated based on the extrapolation error  
\begin{IEEEeqnarray}{rCl}
\mathcal{E}(\omega)&=& 20\log_{10} 
\frac{\sum_{e=1}^{270} | P(\omega,\theta_e,\phi_e) - \hat{P}(\omega,\theta_e,\phi_e)
|}{\sum_{e=1}^{270}|P(\omega,\theta_e,\phi_e)|}, 
\end{IEEEeqnarray}
where $P(\omega,\theta_e,\phi_e)$ and $\hat{P}(\omega,\theta_e,\phi_e)$ 
were the unknown HRTF (ground-truth) and its estimation generated by different methods 
at directions $\{(\theta_e,\phi_e)\}_{e=1}^{270}$, respectively. 
%%%%%%%%%%%%%%%%%%%%%%%%%%%%%%%%%%%%%%%%%%%%%%%%%%%%%%%%%%%%%%%%%%%%%%%%%%%%

%%%%%%%%%%%%%%%%%%%%%%%%%%%%%%%%%%%%%%%%%%%%%%%%%%%%%%%%%%%%%%%%%%%%%%%%%%%%
\subsection{Result: at a single frequency for one subject\label{sec:ext_sphere}}
We presented an extrapolation result in Fig.~\ref{fig:ext_sphere},
which showed the magnitudes of left-ear HRTF of subject 20 at 14.4 kHz,
ground-truth, extrapolations and extrapolation errors of different methods.
%%%%%%%%%%%%%%%%%%%%%%%%%%%%%%%%%%%%%%%%%%%%%%%%%%%%%%%%%%%%%%%%%%%%%%%%%%%%

In the case, the SH method, the NN methods, and the HRTF field method all 
failed to extrapolate the unknown HRTF. Owing to the regularization 
of Eq.~\eqref{eq:regulation}, the SH method \cite{SH2004} was 
unable to accurately represent the known HRTF. 
Similar to Fig.~\ref{fig:int_array}, the NN methods assigned physically invalid values 
to the unknown HRTFs. 
%\textit{The HRTF field method was even }
%%%%%%%%%%%%%%%%%%%%%%%%%%%%%%%%%%%%%%%%%%%%%%%%%%%%%%%%%%%%%%%%%%%%%%%%%%%%

%%%%%%%%%%%%%%%%%%%%%%%%%%%%%%%%%%%%%%%%%%%%%%%%%%%%%%%%%%%%%%%%%%%%%%%%%%%%
As shown in Fig.~\ref{fig:ext_sphere} (g) - (m), without reducing the accuracy 
of representing the known HRTF and without assigning physically invalid values 
to the unknown HRTF, the PINN methods showed better extrapolation results 
than the SH method, the NN methods, and the HRTF field method. 
%%%%%%%%%%%%%%%%%%%%%%%%%%%%%%%%%%%%%%%%%%%%%%%%%%%%%%%%%%%%%%%%%%%%%%%%%%%%
The PINN methods with width $W=U/2$ with depth $L=3$ and $L=4$
achieved the least extrapolation error of $-5.6$ dB and $-6.0$ dB, respectively.
%%%%%%%%%%%%%%%%%%%%%%%%%%%%%%%%%%%%%%%%%%%%%%%%%%%%%%%%%%%%%%%%%%%%%%%%%%%%

%%%%%%%%%%%%%%%%%%%%%%%%%%%%%%%%%%%%%%%%%%%%%%%%%%%%%%%%%%%%%%%%%%%%%%%%%%%%
\subsection{Result: across all frequencies and subjects\label{sec:ext_array}}
Figure~\ref{fig:ext_array} showed the extrapolation errors across all frequencies 
and subjects.
%%%%%%%%%%%%%%%%%%%%%%%%%%%%%%%%%%%%%%%%%%%%%%%%%%%%%%%%%%%%%%%%%%%%%%%%%%%%
Comparing Fig.~\ref{fig:ext_array} with Fig.~\ref{fig:int_array}, we observed 
that at the same frequency the extrapolation errors were larger than the 
interpolation errors  for all methods. 
%%%%%%%%%%%%%%%%%%%%%%%%%%%%%%%%%%%%%%%%%%%%%%%%%%%%%%%%%%%%%%%%%%%%%%%%%%%%

%%%%%%%%%%%%%%%%%%%%%%%%%%%%%%%%%%%%%%%%%%%%%%%%%%%%%%%%%%%%%%%%%%%%%%%%%%%%
Small extrapolation error $\mathcal{E}(\omega)<-20$ dB of the SH method 
could only be achieved at frequency $f=2.1$ kHz. 
%%%%%%%%%%%%%%%%%%%%%%%%%%%%%%%%%%%%%%%%%%%%%%%%%%%%%%%%%%%%%%%%%%%%%%%%%%%%
For frequency above 4.1 kHz, the extrapolation errors of the SH method 
downgraded to be around 0 dB.
%%%%%%%%%%%%%%%%%%%%%%%%%%%%%%%%%%%%%%%%%%%%%%%%%%%%%%%%%%%%%%%%%%%%%%%%%%%%

%%%%%%%%%%%%%%%%%%%%%%%%%%%%%%%%%%%%%%%%%%%%%%%%%%%%%%%%%%%%%%%%%%%%%%%%%%%%
Below 8.2 kHz, the extrapolation errors of the NN methods were comparable to 
those of the PINN methods. 
%%%%%%%%%%%%%%%%%%%%%%%%%%%%%%%%%%%%%%%%%%%%%%%%%%%%%%%%%%%%%%%%%%%%%%%%%%%%
However, above 10.3 kHz, the PINN method extrapolation errors were consistently 
smaller than corresponding NN methods. 
%%%%%%%%%%%%%%%%%%%%%%%%%%%%%%%%%%%%%%%%%%%%%%%%%%%%%%%%%%%%%%%%%%%%%%%%%%%%
At $f=12.3, 14.3$ kHz, the PINN method with depth $L=3$  and 
width $W=U/2$ achieved the smallest average extrapolation errors of 
-4.8 dB and -4.7 dB, respectively.  

The HRTF field method did not exhibit any extrapolation ability. 

\section{Discussion}
\subsection{Experiment results}
From the experiment results shown in Secs. V and VI we saw that all methods' 
performance downgraded with the increment of frequency. 

The SH method's performance was the best in low-frequency range, but its 
performance down-gradation was significant in high-frequency range.

The NN methods achieved  performance comparable to those of 
PINN methods for $f\leq6.2$ kHz. 
Without regularization to the output, the 
NN methods assigned physically invalid values to the 
unknown HRTF in high-frequency range, $f\geq$ 12.3 kHz.

The PINN methods demonstrated the least upsampling errors 
in high-frequency range, $f\geq$ 10.3 kHz, where the upsampling 
was the most challenging. 

The HRTF field method was useful for the interpolation scenario only.

Based on these results, it is recommended to employ the SH method for 
HRTF upsampling in low-frequency range, the NN method in 
mid-frequency range, and the PINN method in high-frequency range, 
respectively.

\subsection{Regularization}
Fundamentally, HRTF upsampling is an ill-conditioned problem. 
To avoid generating physically invalid upsamplings like the NN methods, % shown in 
%Figs.~\ref{fig:int_sphere}, \ref{fig:int_array}, \ref{fig:ext_sphere}, and \ref{fig:ext_array}, 
all upsampling methods have to exploit additional 
information to regularize the upsampling process. 

The SH method~\cite{SH2004} achieved the regularization by constraining the amplitudes 
of the estimated SH coefficients of high-orders. However, the high-order SH coefficients 
are necessary to represent the fine details of HRTF in high-frequency range. 
Thus, constraining their amplitudes will inevitably downgrade the upsampling performance.
%This strategy in essence contradicted with the aim of HRTF upsampling. 

The HRTF field method~\cite{neuralfield}, as well as other learning based methods  
\cite{hrtf_autoencoder1,hrtf_autoencoder2,hrtf_autoencoder3,hrtf_gan1,hrtf_gan2,
film,convolutional}, achieved the regularization through learning implicit 
associations between direction (or ear geometry) and HRTF. 
Their dependence on the training data makes them lacks extrapolation ability as
shown in Figs.~\ref{fig:ext_sphere} and \ref{fig:ext_array}.

\clearpage

The proposed PINN method achieved the regularization through the PDE loss which is 
based on the Helmholtz equation.
The Helmholtz equation dictates that sound pressure is proportional to its
Laplacian
\begin{IEEEeqnarray}{rcl}
\label{eq:diverge}
P=-\frac{1}{(\omega/c)^2} \nabla^2P.
\end{IEEEeqnarray} 
As shown in Figs.~\ref{fig:int_sphere} and  \ref{fig:ext_sphere}, 
this constrained the HRTF upsamplings from taking physically
invalid values like the NN methods.
Furthermore, the Helmholtz equation does not depend on any HRTFs, and thus 
that the PINN method  can extrapolate the unknown HRTFs whose directions
are beyond those of the known HRTFs.

\subsection{PINN width and depth}
To accurately model the training data, researchers tended to design PINN methods to be 
deep and wide~\cite{pinn1,pinn2,pinn3,pinn4}. 
As shown in Figs.~\ref{fig:int_sphere}, 
\ref{fig:int_array}, \ref{fig:ext_sphere}, and \ref{fig:ext_array}, a deep and wide 
PINN method not necessarily achieved the best upsampling results, especially in 
high-frequency range. 
%Worse, it could assign zero, a trivial solution of the Helmholtz equation~\cite{pinn_pp1}, 
%to the unknown HRTFs due to the PDE loss. 
This indicated over-fitting~\cite{pinn_pp1,pinn_pp2,pinn_pp3}. 

To avoid over-fitting, we drew inspiration from the SH decomposition of HRTFs, 
and recommended to set the PINN width as $W=U/2$. 
As shown in Figs.~\ref{fig:int_array} and \ref{fig:ext_array}, with the same depth 
$L$, most cases the PINN methods with width $W=U/2$ performed better than 
the PINN methods with width $W=U$ in high-frequency range $f\geq10.3$ kHz, 
and slightly worse in low-frequency range $f\leq8.2$ kHz. 
%, where the upsampling was the most challenging. 

Figures~\ref{fig:int_array} and \ref{fig:ext_array} also showed that, 
with the same width, the PINN methods of depth $L=3$ performed 
better than the PINN methods of depth $L=2$ in most cases, 
and comparable or slightly better than the PINN methods of depth $L=4$ 
in high-frequency range $f\geq10.3$ kHz.

Further considering that, as shown in Table \ref{tab:parameter},
the number of trainable parameters of the PINN method with depth $L=3$
and width $W=U/2$ was the second least, we recommended to set the PINN method
depth as $L=3$ and width as $W=U/2$, especially in high-frequency range $f\geq10.3$ kHz. 
This could reduce the training time. 
%without compromising the  upsampling
%accuracy.   

%%%%%%%%%%%%%%%%%%%%%%%%%%%%%%%%%%%%%%%%%%%%%%%%%%%%%%%%%%%%%%%%%%%%%%%%%%%%%%%%
Nonetheless, the deeper and wider PINN method with depth $L=4$ and width $W=50$ 
did achieve the least upsampling errors below 6.2 kHz as shown in 
Figs.~\ref{fig:int_array} and \ref{fig:ext_array}. 
This indicated that a deep and wide PINN method may be less susceptible to
over-fitting in low-frequency range  where HRTFs are smoother.
%%%%%%%%%%%%%%%%%%%%%%%%%%%%%%%%%%%%%%%%%%%%%%%%%%%%%%%%%%%%%%%%%%%%%%%%%%%%%%%%

The design of the PINN method, specifically its width and depth, was still empirical.
Further theoretical investigation is needed to  provide 
better guidance on the PINN method design.
This will be one of our future works.

\begin{table}[t]
\centering
\caption{Number of trainable parameters of the PINN methods.}
\begin{tabular}{|l|l|l|}
\hline depth and width & Number of trainable parameters &   $f\leq14.4$ kHz,$U\leq29$      \\
\hline 
$L=2$, $W=U/2$ &   ${U^2}/{4} + {3U} +1$ & $\leq316$ \\
\hline 
$L=2$, $W=U$   &   ${U^2}{} + {6U} +1 $ & $\leq1081$  \\
\hline 
$L=3$, $W=U/2$ &   ${U^2}/{2} + {7U}/2 +1  $ & $\leq556$\\
\hline 
$L=3$, $W=U$   &   ${2U^2}{} + {7U} +1  $  & $\leq2011$\\
\hline 
$L=4$, $W=U/2$ &   ${3U^2}/{4} + {7U}/2 +1  $ & $\leq781$\\
\hline 
$L=4$, $W=U$   &   ${3U^2}{} + {7U} +1  $  & $\leq2911$\\
\hline 
$L=4$, $W=50$   &   7901   & \\
\hline 
\end{tabular}
\label{tab:parameter}
\end{table}
%%%%%%%%%%%%%%%%%%%%%%%%%%%%%%%%%%%%%%%%%%%%%%%%%%%%%%%%%%%%%%%%%%%%%%%%%%
%\subsection{Interpolation}
%From Fig.~\ref{fig:int_array}, we saw that the SH method method works best at the 
%lower frequency range, where the known HRTFs are sufficient for SH coefficient 
%determination. The pure data-driven NN method was the most susceptible to 
%frequency increment, and  achieved larger than $0$ dB interpolation error at $14$ kHz. 
%%%%%%%%%%%%%%%%%%%%%%%%%%%%%%%%%%%%%%%%%%%%%%%%%%%%%%%%%%%%%%%%%%%%%%%%%%

\subsection{HRTF extrapolation}
Comparing Fig.~\ref{fig:ext_array} with Fig.~\ref{fig:int_array}, we 
saw that extrapolation was much more challenging than interpolation.
The -5 dB extrapolation error of the PINN methods for $f\geq10.3$ kHz
was smaller than those of other methods, but may not be enough for accurate 
spatial audio reproduction. 
This indicated that the Helmholtz equation regularization alone
%and the SH informed width 
was insufficient to help the PINN method to generate 
accurate extrapolations in high-frequency range. 

Exploration of additional information, such as human anatomy,  
ear geometries, and cross dataset knowledge,
is necessary to further improve the performance of the PINN method
for HRTF extrapolation.
This will be one of our future works.

\subsection{Limitations}

\textbf{Error metric:}
In this paper, we evaluated the upsampling performance in terms of the MSE of HRTF magnitudes only. 
Error metrics, such as phase error of the upsampling and comparisons of 
HRTF magnitude spectra over frequency between the ground-truth and the upsampling 
in sagittal planes, would also be helpful to assess the upsampling performance. 
Furthermore, it is unclear to which extent the MSE shown in Figs.~5, 6, 8, and 9 
are perceivable.
These limitations will be addressed in a future work.

\textbf{Computational complexity:}
As shown in Table I, the number of trainable parameters of the PINN methods
is small. However, as mentioned in Sec.~\ref{sec:conf} 1), 8$L_\omega$ PINN methods 
are needed to the model the HRTFs of a person for two ears and for $L_\omega$ frequencies. 
The computational complexity will be large. 
The computational complexity can be reduced by building a single PINN method
that models HRTFs for both ears over all $L_\omega$ frequencies like the HRTF 
field method~\cite{neuralfield}. 
This will be one of our future works.

\section{Conclusion}\label{sec:con}
%%%%%%%%%%%%%%%%%%%%%%%%%%%%%%%%%%%%%%%%%%%%%%%%%%%%%%%%%%%%%%%%%%%%%%%%%%%%%%
This paper proposed a PINN method for HRTF upsampling.
%%%%%%%%%%%%%%%%%%%%%%%%%%%%%%%%%%%%%%%%%%%%%%%%%%%%%%%%%%%%%%%%%%%%%%%%%%%%%%
The proposed method exploited the Helmholtz equation,
the governing PDE of acoustic wave propagation, for constraining the 
upsampling process and  generating  physically valid outputs.
%%%%%%%%%%%%%%%%%%%%%%%%%%%%%%%%%%%%%%%%%%%%%%%%%%%%%%%%%%%%%%%%%%%%%%%%%%%%%%
Furthermore, based on the dimensionality of HRTF under SH decomposition and 
the Helmholtz equation, we set the PINN with an appropriate 
width and depth. This helped the PINN method to avoid the over-fitting problem.   
%%%%%%%%%%%%%%%%%%%%%%%%%%%%%%%%%%%%%%%%%%%%%%%%%%%%%%%%%%%%%%%%%%%%%%%%%%%%%%
The Helmholtz equation regularized PINN method with a suitable width and depth
outperformed the SH method, the NN method, and the HRTF field method 
in both interpolation and extrapolation experiments.
%%%%%%%%%%%%%%%%%%%%%%%%%%%%%%%%%%%%%%%%%%%%%%%%%%%%%%%%%%%%%%%%%%%%%%%%%%%%%%

\clearpage

%%%%%%%%%%%%%%%%%%%%%%%%%%%%%%%%%%%%%%%%%%%%%%%%%%%%%%%%%%%%%%%%%%%%%%%%%%%%%%
%\vfill\pagebreak

\end{document}